\documentclass[aps]{revtex4}
\usepackage{amsfonts}
\usepackage{amsmath}
\usepackage{amssymb,epsf}

\begin{document}

\title{Black hole solutions in Gauss-Bonnet-massive gravity \\
in the presence of power-Maxwell field}
\author{S. H. Hendi$^{1,2}$\footnote{
email address: hendi@shirazu.ac.ir}, B. Eslam
Panah$^{1,2}$\footnote{ email address:
behzad.eslampanah@gmail.com} and S. Panahiyan$^{1,3}$\footnote{
email address: shahram.panahiyan@uni-jena.de}} \affiliation{$^{1}$
Physics Department and Biruni Observatory, College of Sciences,
Shiraz University, Shiraz 71454, Iran\\
$^{2}$ Research Institute for Astronomy and Astrophysics of Maragha (RIAAM),
Maragha, Iran\\
$^{3}$ Helmholtz-Institut Jena, Fr\"{o}belstieg 3, Jena D-07743, Germany}

\begin{abstract}
Motivated by recent progresses in the field of massive gravity, the paper at
hand investigates the thermodynamical structure of black holes with three
specific generalizations: i) Gauss-Bonnet gravity which is motivated from
string theory ii) PMI nonlinear electromagnetic field which is motivated
from perspective of the QED correction iii) massive gravity which is
motivated by obtaining the modification of standard general relativity.
The exact solutions of this setup are extracted which are interpreted as
black holes. In addition, thermodynamical quantities of the solutions are
calculated and their critical behavior are studied. It will be shown that
although massive and Gauss-Bonnet gravities are both generalizations in
gravitational sector, they show opposing effects regarding the critical
behavior of the black holes. Furthermore, a periodic effect on number of the
phase transition is reported for variation of the nonlinearity parameter and
it will be shown that for super charged black holes, system is restricted in
a manner that prevents it to reach the critical point and
acquires phase transition. In addition, the effects of geometrical structure
on thermodynamical phase transition will be highlighted.
\end{abstract}

\maketitle

\section{Introduction}

Among the generalizations of Einstein theory of the gravity, Lovelock
generalization has been proven to be special one. This is due to fact that
this theory enjoys the existence of several properties which include: i)
This theory consists curvature-squared terms where interestingly leads to
only up to quadratic order of derivation with respect to metric \cite%
{secondorder,secondorderII}. ii) This theory of gravity enjoys absence of
ghost \cite{ghostfree}. iii) This theory of gravity can solve some of the
shortcomings of general relativity (GR) \cite{shortcomings}. iv) This
gravity arisen from the low-energy limit of heterotic string theory \cite%
{string}. v) This theory may also play an important role in the context of
the braneworld scenario \cite{braneworld}. vi) It is notable that, Lovelock
gravity represents a very interesting scenario to study how higher curvature
corrections to black hole physics substantially change the qualitative
features we know from our experience with black holes in GR.

Gauss-Bonnet (GB) theory was first introduced by Lanczos in ref. \cite%
{Lanczos}, then rediscovered with more details by David Lovelock in ref.
\cite{Lovelock}. The first three terms of Lovelock gravity provide us with
the well known GB gravity which consists of the cosmological constant,
Einstein's tensor and an additional term called the GB term. The first
derivation of black hole solutions in the presence of GB gravity was done by
Boulware and Deser in ref. \cite{secondorder}. The thermodynamical
properties and eikonal instability of these solutions were studied in some
literatures \cite{blackholesGB}. Combination of GB theory with other
theories of gravity such as dilaton \cite{dilaton}, rainbow \cite{rainbowI},
$F(R)$ \cite{F(R)}, and $F(R,G)$ \cite{F(RG)}\ have been done. Charged GB
black hole solutions have been investigated in ref. \cite{Charged}. Later,
exact black hole solutions including their thermodynamical properties have
been explored in ref. \cite{Therodynamics}. Several generalizations of black
holes with nonlinear electrodynamics have also been obtained in ref. \cite%
{nonlinear}.

The effects of GB gravity on different properties of the system have been
investigated in diverse physical contexts, to name a few: i) presentation of
conserved charges of GB-AdS gravity in electric part of the Weyl tensor
which gives a generalization to conformal mass definition \cite{Jatkar}. ii)
investigation of Holographic thermalization in GB gravity through the Wilson
loop and the holographic entanglement entropy \cite{thermalization}. iii)
negative effects of GB parameter on superconductor phase transition through $%
p$-wave phase transition \cite{pwave}. iv) evidences of the effects of GB
gravity of bulk theory on dual conformal field theory of the boundary in
AdS/CFT correspondence \cite{GBAdSCFT,PureGB}. It is worthwhile to mention
that in the presence of GB gravity, gravitons are massless and they
propagate with speed of light. Also, it was shown that on stationary
spacetime, event horizon of black holes in the presence of GB gravity
becomes Killing horizon \cite{Izumi}. Some restrictions regarding the GB
gravity were extracted in ref. \cite{Jose}, in order to avoid violation of
causality. These restrictions pointed out to requirement of an infinite tower
of massive higher-spin states, as it happens in string theory \cite%
{veneziano}, in order to complete the theory.

Recent detection of the gravitational waves from a binary black hole merger
by the LIGO and Virgo collaboration has provided the possibility of
testing the validation of Einstein theory of gravity \cite{Abbott}.
Among the future purposes of developments in detection of gravitational
waves is investigation regarding the possibility of existence of massive
gravitons. The GR and its generalization, GB, include the gravitons as
massless particles \cite{Gupta}. Whereas, in studies conducted in the
context of brane-world gravity, the possible existence of the massive modes
for gravity was pointed out \cite{massiveg}. Massive gravity is a
modification of GR based on the idea of equipping the graviton with mass.
The first ghost-free theory which included massive gravitons in flat space
was proposed by Fierz and Pauli in 1939 \cite{Fierz}. However, it was shown
by Boulware and Deser (BD) in 1972 \cite{Boulware}, that there is an
inevitable ghost instability in generic types of self-interactions for the
massive spin-2 field of Fierz-Pauli in non-flat background. To avoid the
existence of BD ghost in massive gravity, de Rham, Gabadadze and Tolley
(dRGT) in 2011, introduced a set of possible interaction terms to include
massive gravitons and avoid the possibility of BD ghost instability \cite{de
Rham}. In fact, it was shown that a subclass of massive potentials the BD
ghost does not appear, both in dRGT massive gravity \cite{de RhamII} and its
bi-gravity extension \cite{Hinterbichler,Living}.

Black hole solutions and their thermodynamical quantities in the presence of
dRGT massive theory were investigated in ref. \cite{Nieuwenhuizen}. From the
perspective of cosmology and astrophysics, generalization to massive gravity
proved to be significantly fruitful \cite{Saridakis,Babichev}. Among its
achievements in these fields one can point out: i) providing a resolution
for the cosmological constant problem \cite{DvaliGS,DvaliHK} and explaining
the self-acceleration of the universe without introducing the cosmological
constant \cite{Deffayet,DeffayetDG}. ii) providing an equivalent massive
term corresponding to a cosmological constant \cite%
{massivecosmologyA2,GratiaHW,Kobayashi}. iii) resulting into extra
polarization for gravitational waves, and affecting the propagation's speed
of gravitational waves \cite{GW1} and production of gravitational waves
during inflation \cite{GW2,GW3}.

The core stone of producing massive terms in dRGT theory is provided by
employing a reference metric. Accordingly, since the reference metric plays
a key role for constructing the massive theory of gravity \cite{Living},
Vegh introduced a new class of massive theory \cite{Vegh}. Graviton has
similar behavior as lattice in holographic conductor model in this theory.
Also, this theory has specific applications in the gauge/gravity duality
especially in lattice physics which motivate one to use it in other
frameworks as well. Considering Vegh's massive gravity, black hole
solutions, geometrical and their thermodynamical structures were obtained in
ref. \cite{Vegh massiveBH}. Applying this theory to a holographic
superconductor with an implicit periodic potential beyond the probe limit
was done and the results were in agreement with measurements on some
cuprates \cite{Zeng}. The modifications of TOV equation and structure of
neutron stars in the context of this massive gravity were explored and shown
that the maximum mass of neutron stars can be more than three times of the mass
of sun \cite{NSmassive}. Combination of this massive theory with gravity's
rainbow \cite{rainbow}, GB gravity \cite{GBMassive} and nonlinear
electrodynamics were done and the modification in thermodynamical structure
of the black holes were highlighted. Magnetic solutions in the presence
GB-massive (non)linear electrodynamics were studied in ref. \cite{MagMassive}%
, and it was shown that the massive terms significantly affect the
singularity of deficit angle and their possible geometrical phase
transitions. The existence of nonsingular universe in massive gravity's
rainbow has been investigated and pointed out in ref. \cite{nonsingular}.
Entropy spectrum of black holes in the presence massive gravity has been
investigated in ref. \cite{Spectrum}. Black hole as a heat engine within the
framework of massive gravity has been studied in ref. \cite{heat engine}.

Coupling of GR with nonlinear electrodynamics is a subject which
attracted significant attentions because of specific
properties \cite{Heisenberg}. One of the interesting and special
classes of the nonlinear electrodynamic is power-law Maxwell
invariant (PMI) which was introduced by Hassaine and Martinez in 2007
\cite{PMI}. Lagrangian of PMI is an arbitrary power of Maxwell
Lagrangian \cite{PMIII}, where it is invariant under the conformal
transformation $g_{\mu \nu }\rightarrow \Omega ^{2}g_{\mu \nu }$
and $A_{\mu }\rightarrow A_{\mu }$ ($g_{\mu \nu }$ and $A_{\mu }$
are metric tensor and the electromagnetic potential,
respectively). It is notable that, when the power of Maxwell
invariant is unit, this model reduces to linear Maxwell field
\cite{PMIII}. Another interesting property of PMI is related to
its conformal invariancy. Considering power of Maxwell invariant as $%
s=d/4$\ ($s$ and $d$ are power of PMI and dimensions of spacetime,
respectively), one obtains traceless energy-momentum tensor which leads to
conformal invariancy. This conformal invariancy includes Reissner-Nordstr\"{o}m solutions which are inverse square electric field in arbitrary
dimensions \cite{PMI}. The black object solutions coupled to the PMI field
and their thermodynamics were studied and obtained in ref. \cite{PMIresult}.
Also, holographic superconductors in the presence of PMI field were
investigated in ref. \cite{PMIHS}.

In this paper, we intend to generalize Einstein gravity action by adding GB,
massive and nonlinear source to Lagrangian and obtain exact GB-Massive black
holes in the presence of PMI source.

\section{Charged Black hole solutions in GB-Massive gravity \label{GBMassive}}

The $d$-dimensional action of GB-massive gravity in the presence of negative
cosmological constant and power Maxwell field can be written as%
\begin{equation}
I=-\frac{1}{16\pi }\int d^{d}x\sqrt{-g}\left[ R-2\Lambda +L(F)+\alpha
L_{GB}-m^{2}\sum_{i}^{4}c_{i}{}U_{i}(g,\psi )\right] ,  \label{Action}
\end{equation}%
where $R$\ is the scalar curvature, $L(F)=\left( -F\right) ^{s}$\ is the
Lagrangian of power Maxwell theory and also $F=F_{\mu \nu }F^{\mu \nu }$\
(in which $F_{\mu \nu }=\partial _{\mu }A_{\nu }-\partial _{\nu }A_{\mu }$)
is the Maxwell invariant. $\alpha $\ and $L_{GB}=R_{\mu \nu \gamma \delta
}R^{\mu \nu \gamma \delta }-4R_{\mu \nu }R^{\mu \nu }+R^{2}$ are,
respectively, the coefficient and the Lagrangian of GB gravity. $\psi $ is a
fixed symmetric fiducial metric tensor. In Eq. (\ref{Action}), $c_{i}$ are
constants and $U_{i}$'s are symmetric polynomials of the eigenvalues of the $%
d\times d$\ matrix $K_{\nu }^{\mu }=\sqrt{g^{\mu \alpha }\psi _{\alpha \nu }}
$
\begin{eqnarray}
U_{1} &=&\left[ K\right] ,\ \ \ \ U_{2}=\left[ K\right] ^{2}-\left[ K^{2}%
\right] ,\ \ \ \ U_{3}=\left[ K\right] ^{3}-3\left[ K\right] \left[ K^{2}%
\right] +2\left[ K^{3}\right] ,  \notag \\
U_{4} &=&\left[ K\right] ^{4}-6\left[ K^{2}\right] \left[ K\right] ^{2}+8%
\left[ K^{3}\right] \left[ K\right] +3\left[ K^{2}\right] ^{2}-6\left[ K^{4}%
\right] ,  \notag \\
U_{5} &=&\left[ K\right] ^{5}-10\left[ K^{2}\right] \left[ K\right] ^{3}+20%
\left[ K^{3}\right] \left[ K\right] ^{2}-20\left[ K^{2}\right] \left[ K^{3}%
\right] +15\left[ K\right] \left[ K^{2}\right] ^{2}  \notag \\
&&-30\left[ K\right] \left[ K^{4}\right] +24\left[ K^{5}\right] ,  \notag \\
&&...~~~.
\end{eqnarray}

Here, technically, it is not easy to study the solutions with all
possible $U_{i}$'s in an arbitrary $d-$dimensions (for $i>d-2$, 
$U_{i}=0$). GB gravity is the general Lagrangian for $5$ and $6$
dimensions with Einstein assumptions of general relativity, and it
is clear that $U_{i}=0,($i>4$)$ for $d=6$. Although we generalize
the solutions to higher dimensions, for the sake of simplicity, we
keep the massive effects up to $U_{4}$. Obviously, if we regard
other higher curvature terms of Lovelock theory, it is logical to
consider other $U_{i}$.

We can extract the equations of motion by variation of the action (\ref%
{Action}) with respect to the metric tensor $g_{\mu \nu }$\ and the Faraday
tensor $F_{\mu \nu }$, so we have
\begin{equation}
R_{\mu \nu }-\frac{1}{2}Rg_{\mu \nu }+\Lambda g_{\mu \nu }+H_{\mu \nu
}-2sF_{\mu \lambda }F_{\nu }^{\lambda }\left( -F\right) ^{s-1}-\frac{1}{2}%
g_{\mu \nu }\left( -F\right) ^{s}+m^{2}\chi _{\mu \nu }=0,
\label{Field equation}
\end{equation}%
\begin{equation}
\nabla _{\mu }\left(F^{s-1} F^{\mu \nu }\right) =0,  \label{Maxwell equation}
\end{equation}%
where $\chi _{\mu \nu }$\ and $H_{\mu \nu }$\ are in the following forms
\begin{eqnarray}
\chi _{\mu \nu } &=&-\frac{c_{1}}{2}\left( U_{1}g_{\mu \nu }-K_{\mu \nu
}\right) -\frac{c_{2}}{2}\left( U_{2}g_{\mu \nu }-2U_{1}K_{\mu \nu }+2K_{\mu
\nu }^{2}\right) -\frac{c_{3}}{2}\left( U_{3}g_{\mu \nu }-3U_{2}K_{\mu \nu
}+6U_{1}K_{\mu \nu }^{2}-6K_{\mu \nu }^{3}\right)  \notag \\
&&  \notag \\
&&-\frac{c_{4}}{2}\left( U_{4}g_{\mu \nu }-4U_{3}K_{\mu \nu }+12U_{2}K_{\mu
\nu }^{2}-24U_{1}K_{\mu \nu }^{3}+24K_{\mu \nu }^{4}\right) , \\
&&  \notag \\
H_{\mu \nu } &=&-\frac{\alpha }{2}\left( 8R^{\rho \sigma }R_{\mu \rho \nu
\sigma }-4R_{\mu }^{\rho \sigma \lambda }R_{\nu \rho \sigma \lambda
}-4RR_{\mu \nu }+8R_{\mu \lambda }R_{\nu }^{\lambda }+g_{\mu \nu
}L_{GB}\right) .
\end{eqnarray}

We consider a topological $d$-dimensional static spacetime as
\begin{equation}
ds^{2}=-f(r)dt^{2}+f^{-1}(r)dr^{2}+r^{2}h_{ij}dx_{i}dx_{j},\
i,j=1,2,3,...,n~,  \label{Metric}
\end{equation}
where $f(r)$ is the metric function and $h_{ij}dx_{i}dx_{j}$ is the line
element of a ($d-2$)-dimensional space with constant curvature $(d-2)(d-3)k$
and volume $V_{d-2}$. We should note that the constant $k$ indicates that
the boundary of $t=constant$ and $r=constant$ can be elliptic ($k=1$), flat (%
$k=0$) or hyperbolic ($k=-1$) curvature hypersurface.

In order to find the metric function, analytically, we should consider a
suitable reference metric at the first step \cite{Vegh,Vegh massiveBH}
\begin{equation}
\psi _{\mu \nu }=diag(0,0,c^{2}h_{ij}),  \label{f11}
\end{equation}%
where $c$ is positive constant. Using the Eq. (\ref{f11}), $U_{i}$'s be
written as \cite{Vegh,Vegh massiveBH}
\begin{equation}
U_{j}=\frac{c^{j}}{r^{j}}\Pi _{k=2}^{j+1}d_{k},
\end{equation}%
where $d_{k}=d-k$. Using the electrical gauge potential ansatz, $A_{\mu
}=h(r)\delta _{\mu }^{0}$, with Maxwell equation (\ref{Maxwell equation}),
we obtain
\begin{equation}
rh{^{\prime \prime }}(r) ( 2s-1) +d_{2}h{^{\prime }}(r) =0.
\end{equation}

Using the above equation we can extract the electrical gauge potential as
\begin{equation}
h(r)=\left\{
\begin{array}{cc}
q, & s=0,1/2 \\ \\
-q\ln \left( \frac{r}{l}\right) , & s=\frac{d_{1}}{2} \\ \\
-q\sqrt{\frac{d_{2}}{2d_{3}}}r^{\frac{2s-d_{1}}{2s-1}}, & otherwise%
\end{array}%
\right. ,
\end{equation}%
where $q$\ is an integration constant which is related to the electric
charge. Physically, the electromagnetic gauge
potential should be finite at infinity, therefore, one should impose
following restriction on the nonlinearity parameter
\begin{equation}
\frac{2s-d_{1}}{2s-1}<0.
\end{equation}

Applying the above equation, we find a restriction on the range of $s$ as
\begin{equation}
\frac{1}{2}<s<\frac{d_{1}}{2}.
\end{equation}

To find metric function $f(r)$ for the metric (\ref{Metric}), one may use
any components of Eq. (\ref{Field equation}). For this purpose, we can use
different components ($tt$ and $x_{1}x_{1}$) of Eq. (\ref{Field equation})
which can be written as
\begin{eqnarray}
tt-component &=&\frac{d_{2}d_{3}f}{2r^{2d_{2}}}\{d_{4}d_{5}\left[
m^{2}c^{4}c_{4}+\alpha f^{2}\right] r^{2d_{4}}+d_{4}\left[
m^{2}c^{3}c_{3}+2\alpha ff{^{\prime }}\right] r^{2d_{7/2}}+\left[
m^{2}c^{2}c_{2}-f\right] r^{2d_{3}}  \notag \\
&&+\frac{\left[ m^{2}cc_{1}-f{^{\prime }}\right] r^{2d_{5/2}}}{d_{3}}-\frac{%
2\Lambda r^{2d_{2}}}{d_{2}d_{3}}-r^{2d_{3}}k\left( \alpha d_{4}d_{5}\left[
2f-k\right] r^{-2}+2\alpha d_{4}f^{\prime }r^{-1}-1\right) \}+\Phi _{1},
\label{tteq} \\
&&  \notag \\
x_{1}x_{1} -component&=&\frac{-d_{3}d_{4}}{2r^{2d_{3}}}\{d_{5}d_{6}\left[
m^{2}c^{4}c_{4}+\alpha f^{2}\right] r^{2d_{4}}+d_{5}\left[
m^{2}c^{3}c_{3}+4\alpha ff{^{\prime }}\right] r^{2d_{7/2}}-\frac{\left(
2\Lambda +f{^{\prime \prime }}\right) r^{2d_{2}}}{d_{3}d_{4}}  \notag \\
&&+\left[ m^{2}c^{2}c_{2}+2\alpha f{^{\prime 2}+2\alpha f}f{^{\prime \prime }%
}-f\right] r^{2d_{3}}+\frac{\left[ m^{2}cc_{1}-2f{^{\prime }}\right]
r^{2d_{5/2}}}{d_{4}}  \notag \\
&&-r^{2d_{3}}k\left( \alpha d_{5}d_{6}\left[ 2f-k\right] r^{-2}+4\alpha
d_{5}f^{\prime }r^{-1}+2\alpha f^{\prime \prime }-1\right) \}+\Phi _{2}.
\label{x1x1eq}
\end{eqnarray}%
where $\Phi _{1}$ and $\Phi _{2}$ are, respectively,
\begin{eqnarray}
\Phi _{1} &=&\left\{
\begin{array}{cc}
0 & s=0,1/2 \\ \\
-\left( 2\right) ^{\frac{d_{3}}{2}}d_{2}f\left( \frac{q}{r}\right) ^{d_{1}},
& s=\frac{d_{1}}{2}  \\  \\
-\frac{\left( 2s-1\right) f}{2}\left( \frac{d_{2}q^{2}\left( 2s-d_{1}\right)
^{2}}{d_{3}\left( 2s-1\right) ^{2}}r^{\frac{-2d_{2}}{2s-1}}\right) ^{s}, &
otherwise%
\end{array}%
\right. , \\ \\
&&  \notag \\ \\
\Phi _{2} &=&\left\{
\begin{array}{cc}
0 & s=0,1/2 \\  \\
-2\left( 2\right) ^{\frac{d_{5}}{2}}\frac{q^{d_{1}}}{r^{d_{3}}}, & s=\frac{%
d_{1}}{2} \\  \\
-\frac{r^{2}}{2}\left( \frac{d_{2}q^{2}\left( 2s-d_{1}\right) ^{2}}{%
d_{3}\left( 2s-1\right) ^{2}}r^{\frac{-2d_{2}}{2s-1}}\right) ^{s}, &
otherwise%
\end{array}%
\right. .
\end{eqnarray}

Now, by using the Eqs. (\ref{tteq}) and (\ref{x1x1eq}), we can obtain the
metric function as
\begin{eqnarray}
f\left( r\right) &=&k+\frac{r^{2}}{2\alpha d_{3}d_{4}}\left\{ 1-\sqrt{1+%
\frac{4\alpha d_{3}d_{4}}{d_{2}}\left[ \frac{2\Lambda }{d_{1}}+\frac{%
d_{2}m_{0}}{r^{d_{1}}}+\Gamma \left( r\right) +\Upsilon \right] }\right\} ,
\label{f(r)} \\
&&  \notag \\
\Upsilon &=&-m^{2}d_{2}\left[ \frac{d_{3}d_{4}c^{4}c_{4}}{r^{4}}+\frac{%
d_{3}c^{3}c_{3}}{r^{3}}+\frac{c^{2}c_{2}}{r^{2}}+\frac{cc_{1}}{d_{2}r}\right]
, \\
&&  \notag \\
\Gamma \left( r\right) &=&\left\{
\begin{array}{cc}
0 & s=0,1/2 \\ \\
2^{d_{1}/2}d_{2}\left( \frac{q}{r}\right) ^{d_{1}}\ln \left( \frac{r}{l}%
\right) , & s=\frac{d_{1}}{2} \\  \\  \frac{\left( 2s-1\right)
^{2}}{2s-d_{1}}\left( \frac{d_{2}q^{2}\left(
2s-d_{1}\right) ^{2}}{d_{3}\left( 2s-1\right) ^{2}}r^{\frac{-2d_{2}}{2s-1}%
}\right) ^{s} & otherwise%
\end{array}%
\right. .
\end{eqnarray}

In the above equation, $m_{0}$\ is integration constants which is related to
the total mass of the black hole. It is notable that, obtained solutions (%
\ref{f(r)}), satisfy all components of Eq. (\ref{Field equation}).

Here, we want to study the existence of the singularity for obtained
solutions. To find that, we investigate the Kretschmann scalar.
Straightforward calculations show that the Kretschmann scalar for small
values leads to
\begin{equation}
{\lim_{r\longrightarrow 0}}R_{\alpha \beta \gamma \delta }R^{\alpha \beta
\gamma \delta }=\infty ,  \label{RRzero}
\end{equation}%
so, there is an essential singularity located at the origin ($r=0$). In
order to study asymptotical behavior of the metric solutions, one can
investigate the Kretschmann scalar for large values of radial coordinate. We
find it as
\begin{equation}
{\lim_{r\longrightarrow \infty }}R_{\alpha \beta \gamma \delta }R^{\alpha
\beta \gamma \delta }=\frac{\left[ d_{-1}d_{1}d_{2}+\left( 4\Lambda
d_{-1}+d_{1}\right) \alpha d_{3}d_{4}\right] d_{2}-\left(
d_{-1}d_{2}+d_{3}d_{4}\alpha \right) \sqrt{d_{1}d_{2}\left[
8d_{3}d_{4}\alpha \Lambda +d_{1}d_{2}\right] }}{d_{1}d_{2}d_{3}^{2}d_{4}^{2}%
\alpha ^{2}}.  \label{RRlarge}
\end{equation}

Also for small values of GB parameter it will yield
\begin{equation}
{\lim_{\alpha \longrightarrow 0 }}R_{\alpha \beta \gamma \delta }R^{\alpha
\beta \gamma \delta }=\frac{8d_{-1}}{d_{1}^{2}d_{2}}\Lambda ^{2}-\frac{4}{%
d_{1}d_{2}}\Lambda +O\left( \alpha \right) .  \label{RRexpand}
\end{equation}

The above equations (Eqs. (\ref{RRlarge}) and (\ref{RRexpand})), confirm
that asymptotical behavior of the solutions is (A)dS with an effective
cosmological constant $\Lambda _{eff}=\Lambda _{eff}(\Lambda ,\alpha )$.
Therefore, the solution (\ref{f(r)}), describes a $d$-dimensional
asymptotically (A)dS topological black hole with a negative, zero or
positive constant curvature hypersurface in GB-massive gravity in the
presence of power Maxwell field.

Also, in the absence of massive parameter ($m=0$), the solution (\ref{f(r)})
reduces to the obtained metric function for GB gravity in the presence power
Maxwell field \cite{GBPMI}
\begin{equation}
f\left( r\right) =k+\frac{r^{2}}{2\alpha d_{3}d_{4}}\left\{ 1-\sqrt{1+\frac{%
4\alpha d_{3}d_{4}}{d_{2}}\left[ \frac{2\Lambda }{d_{1}}+\frac{d_{2}m_{0}}{%
r^{d_{1}}}+\Gamma \left( r\right) +\Upsilon \right] }\right\} .
\end{equation}

\begin{figure}[tbp]
\epsfxsize=8cm \centerline{\epsffile{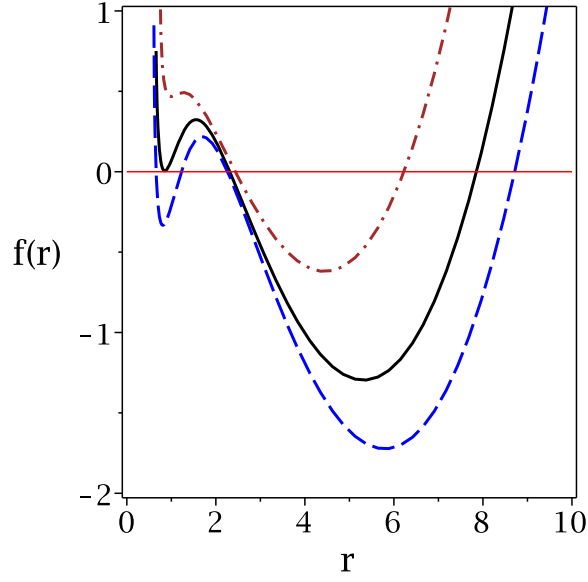}}
\caption{$f(r)$ versus $r$ for $\Lambda =-1$, $q=1$, $\protect\alpha =1.4$, $%
m=2$, $c_{1}=-0.8$, $c_{2}=-0.5$, $c_{3}=1$, $c_{4}=-0.8$, $s=0.7$, $%
m_{0}=0.4$, $k=1$, $d=5$, $c=1.00$ (dashed-dotted line), $c=1.19$ (continues
line) and $c=1.30 $ (dashed line).}
\label{Fig1}
\end{figure}

\begin{figure}[tbp]
$%
\begin{array}{cc}
\epsfxsize=7cm \epsffile{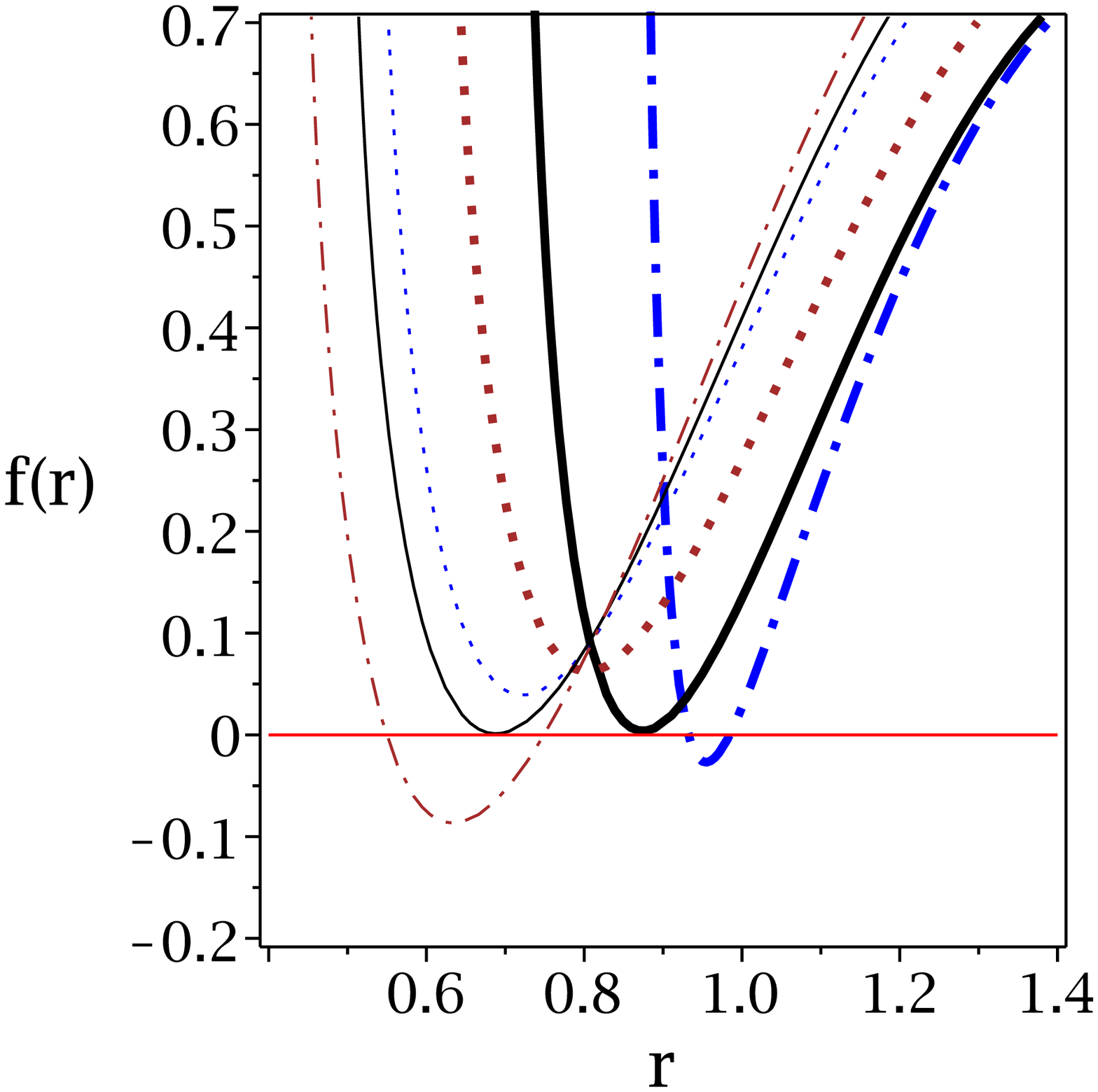} & \epsfxsize=7cm %
\epsffile{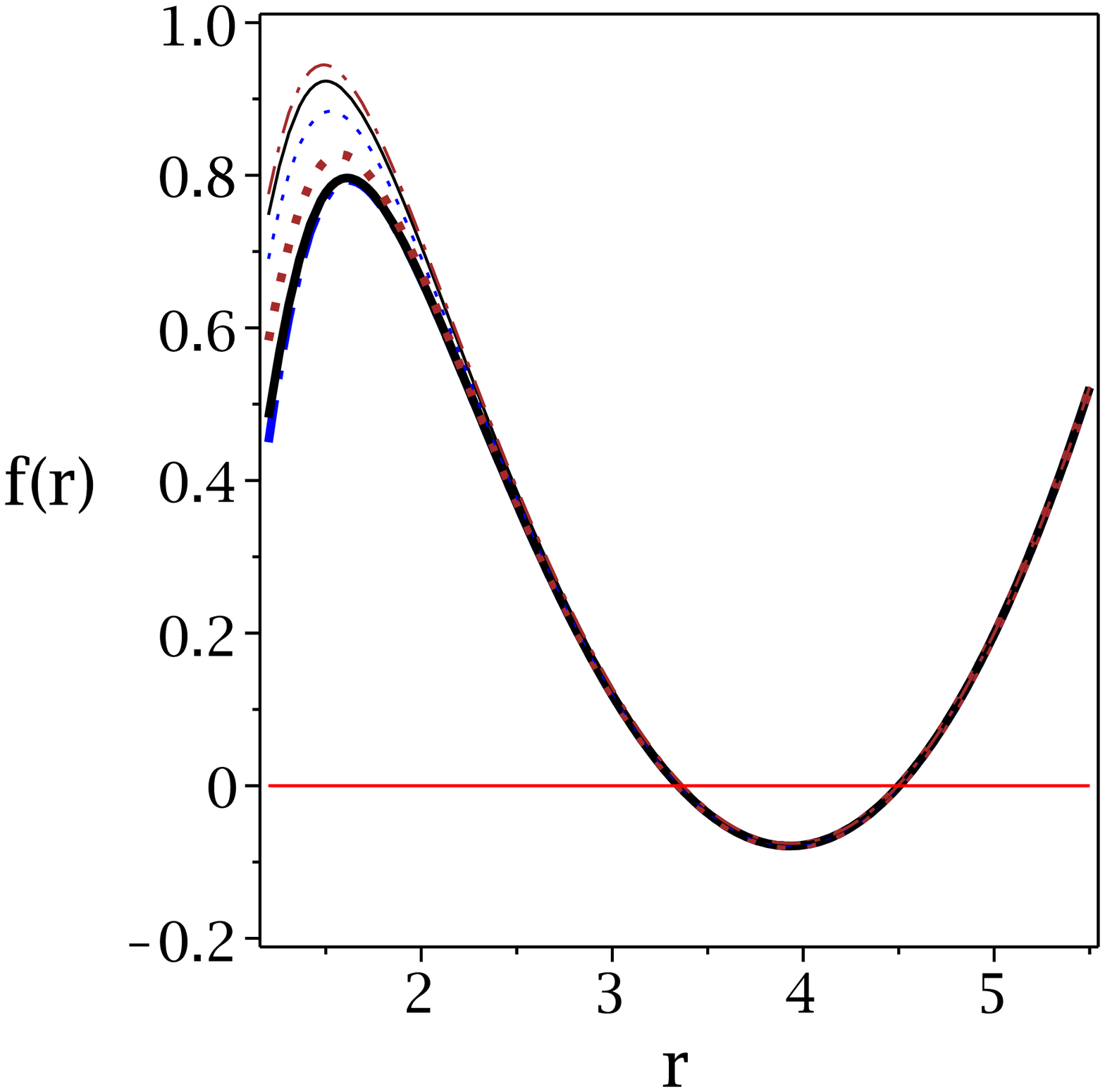}%
\end{array}
$%
\caption{$f(r)$ versus $r$ for $\Lambda =-1.2$, $q=1$, $\protect\alpha =0.4$%
, $m=2$, $c=1.1$, $c_{1}=-0.8$, $c_{2}=-0.35$, $c_{3}=1$, $c_{4}=-0.8$, $%
m_{0}=0.4$, $k=1$, $d=5$, $s=0.55$ (bold dashed-dotted line), $s=0.68$ (bold
continues line), $s=0.80$ (bold dotted line), $s=0.9$ (dotted line), $s=0.93$
(continues line) and $s=0.98$ (dashed-dotted line).}
\label{Fig2}
\end{figure}

The existence of at least one horizon for the obtained metric function shows that
these solutions may be interpreted as black hole solutions. Considering the
specific values of different parameters of massive gravity, the metric
function will have two different cases; i) the usual behavior similar to
GB-Maxwell gravity (two horizons, or one extreme horizon, or no horizon
(naked singularity), see Figs. \ref{Fig1} and \ref{Fig2}, for more details).
ii) the existence of more than two roots for the metric function. In other
words, we encounter with interesting solutions which have two normal and one
extreme (outer) horizons or four horizons (see Figs.\ref{Fig1} and \ref{Fig2}%
, for more details). In other words, the massive parameters affect the
number of roots of the metric function.

%%%%%%%%%%%%%%%%%%%%%%%%%%%%%%%%%%%%%%%%%%%%%%%%%%%%%%%%%%%%%%%
%%%%%%%%%%%%%%%%%%%%%%%%%%%%%%%%%%%%%%%%%%%%%%%%%%%%%%%%%%%%%%%%
%%%%%%%%%%%%%%%%%%%%%%%%%%%%%%%%%%%%%%%%%%%%%%%%%%%%%%%%%%%%%%%

%%%%%%%%%%%%%%%%%%%%%%%%%%%%%%%%%%%%%%%%%%%%%%%%%%%%%%%%%%%%%%%
%%%%%%%%%%%%%%%%%%%%%%%%%%%%%%%%%%%%%%%%%%%%%%%%%%%%%%%%%%%%%%%%

\section{Thermodynamics \label{Thermo}}

In order to investigate the first law of thermodynamics, first, we calculate
the conserved and thermodynamics quantities of the static black hole
solutions in $d$-dimensional GB-massive context, and then by using these
quantities we examine the first law of thermodynamics.

Applying the definition of surface gravity at the outer horizon $r_{+}$, we
obtain the Hawking temperature of this black hole in the following form
\begin{eqnarray}
T &=&\frac{1}{4\pi r_{+}\left[ r_{+}^{2}+2\alpha kd_{3}d_{4}\right] }\left\{
\Xi _{1}+d_{3}k\left( \alpha d_{4}d_{5}k+r_{+}^{2}\right) -\frac{r_{+}^{4}}{%
d_{2}}\left( 2\Lambda +\Xi _{2}\right) \right\} ,  \label{TotalT} \\
\end{eqnarray}
where $\Xi _{1}$ and $\Xi _{2}$ are, respectively, the massive and PMI
effects on the temperature with the following explicit forms
\begin{eqnarray}
\Xi _{1} &=&m^{2}\left(
cc_{1}r_{+}^{3}+d_{3}c^{2}c_{2}r_{+}^{2}+d_{3}d_{4}c^{3}c_{3}r_{+}+d_{3}d_{4}d_{5}c^{4}c_{4}\right) ,
\\
&&  \notag \\
\Xi _{2} &=&\left\{
\begin{array}{cc}
d_{2}2^{\frac{d_{1}}{2}}\left( \frac{q}{r_{+}}\right) ^{d_{1}}, & s=\frac{%
d_{1}}{2} \\
\left( 1-2s\right) \left( \frac{d_{2}q^{2}\left( d_{1}-2s\right) ^{2}}{%
d_{3}\left( 2s-1\right) ^{2}}r_{+}^{\frac{-2d_{2}}{2s-1}}\right) ^{s}, &
otherwise%
\end{array}%
\right. .
\end{eqnarray}

We use the flux of the electromagnetic field at infinity to calculate the
electric charge of the black hole, yielding
\begin{equation}
Q=\left\{
\begin{array}{cc}
-\frac{2^{d_{9}/2}d_{1}}{\pi }q^{d_{2}}, & s=\frac{d_{1}}{2} \\
\frac{s\left( 2s-1\right) q^{2s-1}}{2^{7/2}\pi \sqrt{\frac{d_{2}}{d_{3}}}%
\left( d_{1}-2s\right) }\left( \frac{d_{2}\left( d_{1}-2s\right) ^{2}}{%
d_{3}\left( 2s-1\right) ^{2}}\right) ^{s}, & otherwise%
\end{array}%
\right. .  \label{TotalQ}
\end{equation}

Next, we use the definition introduced in ref. \cite{blackholesGB} for
calculating the electric potential, $U$, as
\begin{equation}
U=A_{\mu }\chi ^{\mu }\left\vert _{r\rightarrow reference }\right. -A_{\mu
}\chi ^{\mu }\left\vert _{r\rightarrow r_{+}}\right. =\left\{
\begin{array}{cc}
q\ln \left( \frac{r_{+}}{l}\right) & s=\frac{d_{1}}{2} \\
&  \\
q\sqrt{\frac{d_{2}}{2d_{3}}}r_{+}^{\frac{2s-d_{1}}{2s-1}}, & otherwise%
\end{array}%
\right. .  \label{TotalU}
\end{equation}

Due to generalization of Einstein gravity to GB gravity, one can use Wald's
formula to obtain the entropy of the black holes \cite{blackholesGB}. This
leads to
\begin{equation}
S=\frac{V_{d_{2}}}{4}r_{+}^{d_{2}}\left( 1+\frac{2d_{2}d_{3}}{r_{+}^{2}}%
k\alpha \right) .  \label{TotalS}
\end{equation}

It is notable that, the above equation shows that the area law is violated
for the GB black holes with non-flat horizons.

Another important conserved quantity is the total mass of black holes. For
finding it, one can use the Hamiltonian approach which results into
\begin{equation}
M=\frac{V_{d_{2}}\ d_{2}\ m_{0}}{16\pi }.  \label{TotalM}
\end{equation}

Now, we define the intensive parameters conjugating to $S$\ and $Q$. These
quantities are the temperature and the electric potential with the following
form
\begin{equation}
T=\left( \frac{\partial M}{\partial S}\right) _{Q}\ \ \ \ \ \ \ \&\ \ \ \ \
\ \ \ U=\left( \frac{\partial M}{\partial Q}\right) _{S}.  \label{TU}
\end{equation}

The above results coincide with Eqs. (\ref{TotalT}) and (\ref{TotalU})
and, therefore, we find that these conserved and thermodynamic quantities
satisfy the first law of black hole thermodynamics as
\begin{equation}
dM=TdS+UdQ.
\end{equation}

\section{Heat capacity \label{Stability}}

The heat capacity is one of the thermodynamical quantities carrying crucial
information regarding thermodynamical structure of the black holes. There
are three specific information that are of interest regarding heat capacity:
i) the discontinuities of this quantity mark the possible thermal phase
transitions that system can undergo. ii) the sign of it determines whether
the system is thermally stable or not. In other words, the positivity
corresponds to thermal stability while the opposite shows instability. iii)
the roots of this quantity are also of interest since it may yield the
possible changes between stable/instable states or bound point. Due to these
important points, this section and the following one are dedicated to
calculation of the heat capacity of the solutions and investigation of
thermal structure of the black holes using such quantity. We will show that
by using this quantity alongside of the temperature, we can draw a picture
regarding the possible thermodynamical phase structures of these black holes
and stability/instability that these black holes could enjoy/suffer.

One can use following relation for calculating the heat capacity
\begin{equation}
C_{Q}=\frac{T}{\left( \frac{\partial ^{2}M}{\partial S^{2}}\right) _{Q}}%
=T(r_{+})\left( \frac{\partial S(r_{+})}{\partial T(r_{+})}\right) _{Q}=T%
\frac{\left( \frac{\partial S}{\partial r_{+}}\right) _{Q}}{\left( \frac{%
\partial T}{\partial r_{+}}\right) _{Q}}.  \label{CQ}
\end{equation}

\subsection{Special case: $s=\frac{d_{1}}{2}$:}

Considering Eqs. (\ref{TotalT}) and (\ref{TotalS}) for the case $s= \frac{%
d_{1}}{2}$, we can obtain the heat capacity in the following form%
\begin{equation}
C_{Q}=\frac{B_{1}}{C_{1}},  \label{HeatI}
\end{equation}%
in which $B_{1}$\ and $C_{1}$\ are
\begin{eqnarray}
B_{1} &=&-\frac{d_{2}^{2}r_{+}^{d_{2}}}{2}\left( d_{3}d_{4}\alpha k\left[
d_{3}d_{4}\alpha k+r_{+}^{2}\right] +\frac{r_{+}^{4}}{4}\right) \left[ d_{3}k%
\left[ d_{4}d_{5}\alpha k+r_{+}^{2}\right] +\Xi _{1_{+}}-\frac{r_{+}^{4}}{%
d_{2}}\left( 2\Lambda +2^{\frac{d_{1}}{2}}\left( \frac{q}{r_{+}}\right)
^{d_{1}}\right) \right] , \\
&&  \notag \\
C_{1} &=&r_{+}^{2}\left\{ d_{2}d_{3}d_{4}\alpha k^{2}\left[
d_{3}d_{4}d_{5}\alpha k+\frac{d_{9}r_{+}^{2}}{2}\right] +k\left(
d_{3}d_{4}\alpha r_{+}^{4}\left[ 6\Lambda -d_{2}d_{4}2^{\frac{d_{1}}{2}%
}\left( \frac{q}{r_{+}}\right) ^{d_{1}}+\frac{d_{2}}{2d_{4}\alpha }\right]
\right. \right.  \notag \\
&&\left. -m^{2}d_{2}d_{3}d_{4}\alpha \left[
2cc_{1}r_{+}^{3}+d_{3}c^{2}c_{2}r_{+}^{2}+d_{3}d_{4}d_{5}c^{4}c_{4}\right]
\right) +\frac{m^{2}d_{2}d_{3}r_{+}^{2}}{2}\left[
c^{2}c_{2}r_{+}^{2}+2d_{4}c^{3}c_{3}r_{+}+3d_{4}d_{5}c^{4}c_{4}\right]
\notag \\
&&\left. +r_{+}^{6}\left( \Lambda -\frac{2^{\frac{d_{1}}{2}}d_{2}^{2}}{2}%
\left( \frac{q}{r_{+}}\right) ^{d_{1}}\right) \right\} ,
\end{eqnarray}%
and also $\Xi _{1_{+}}=\Xi _{1_{\left\vert _{r=r_{+}}\right. }}$.

\subsection{general case: $s\neq \frac{d_{1}}{2}$:}

Considering Eqs. (\ref{TotalT}) and (\ref{TotalS}) for case $s\neq \frac{d_{1}}{%
2}$, we can obtain the heat capacity 
\begin{equation}
C_{Q}=\frac{B_{2}}{C_{2}},  \label{HeatII}
\end{equation}%
where $B_{2}$\ and $C_{2}$\ are
\begin{eqnarray}
B_{2} &=&-\frac{d_{2}^{2}r_{+}^{d_{2}}}{2}\left( d_{3}d_{4}\alpha k\left[
d_{3}d_{4}\alpha k+r_{+}^{2}\right] +\frac{r_{+}^{4}}{4}\right) \left[ d_{3}k%
\left[ d_{4}d_{5}\alpha k+r_{+}^{2}\right] +\Xi _{1_{+}}-\frac{r_{+}^{4}}{%
d_{2}}\left( 2\Lambda +\frac{\Xi _{2_{+}}}{\left( 1-2s\right) }\right) %
\right] , \\
&&  \notag \\
C_{2} &=&r_{+}^{2}\left\{ d_{2}d_{3}d_{4}\alpha k^{2}\left[
d_{3}d_{4}d_{5}\alpha k+\frac{d_{9}r_{+}^{2}}{2}\right] +k\left(
d_{3}d_{4}\alpha r_{+}^{4}\left[ 6\Lambda -\frac{\left( 2ds-10s+3\right) \Xi
_{2_{+}}}{\left( 1-2s\right) }+\frac{d_{2}}{2d_{4}\alpha }\right] \right.
\right.  \notag \\
&&\left. -m^{2}d_{2}d_{3}d_{4}\alpha \left[
2cc_{1}r_{+}^{3}+d_{3}c^{2}c_{2}r_{+}^{2}+d_{3}d_{4}d_{5}c^{4}c_{4}\right]
\right) +\frac{m^{2}d_{2}d_{3}r_{+}^{2}}{2}\left[
c^{2}c_{2}r_{+}^{2}+2d_{4}c^{3}c_{3}r_{+}+3d_{4}d_{5}c^{4}c_{4}\right]
\notag \\
&&\left. +r_{+}^{6}\left( \Lambda -\frac{\left( 2ds-6s+1\right) \Xi _{2_{+}}%
}{2\left( 2s-1\right) ^{2}}\right) \right\} ,
\end{eqnarray}%
and also $\Xi _{2_{+}}=\Xi _{2_{\left\vert _{r=r_{+}}\right. }}$.

\section{Results, discussion and conclusion}

\begin{figure}[tbp]
$%
\begin{array}{cc}
\epsfxsize=7cm \epsffile{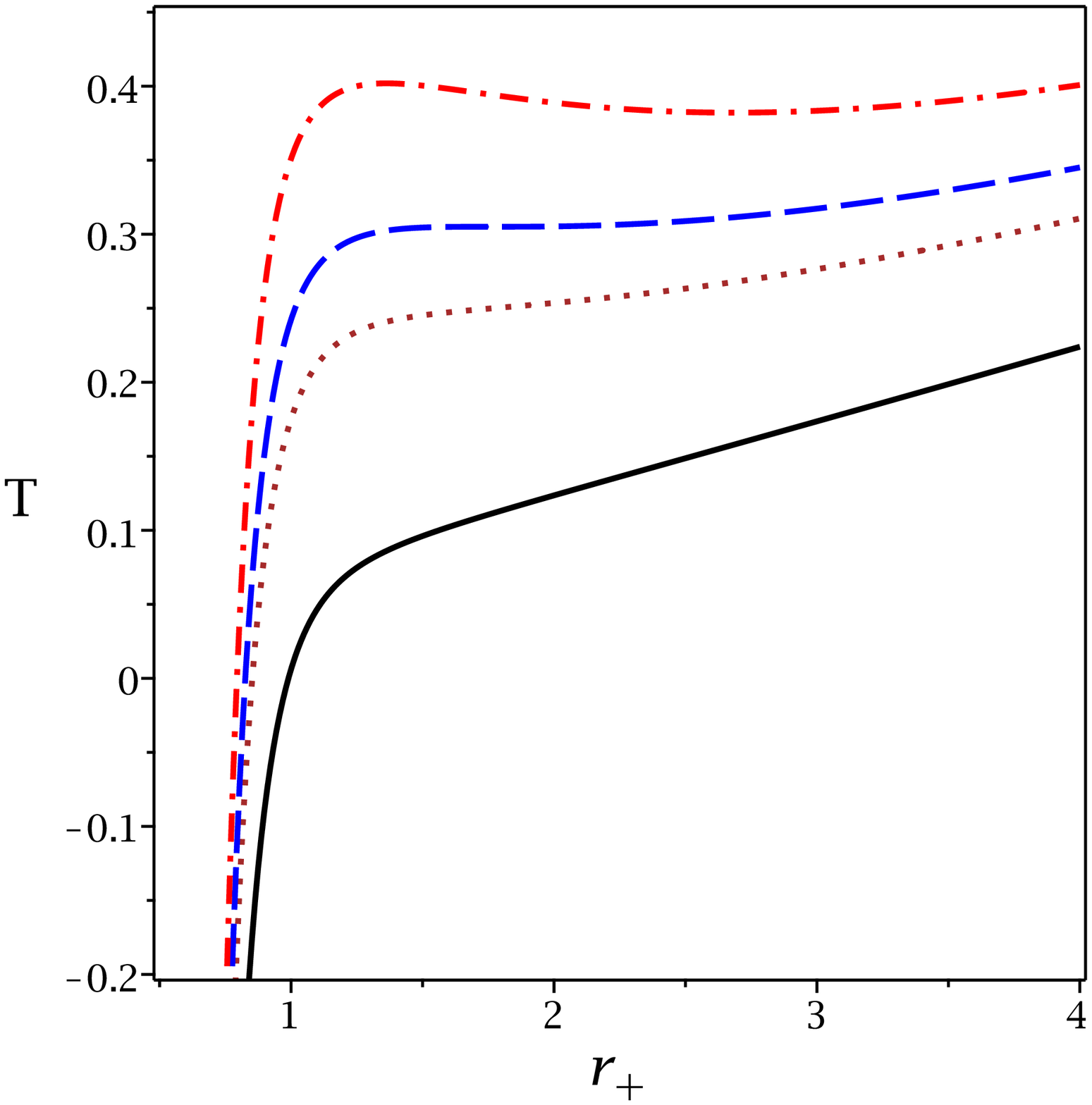} & \epsfxsize=7cm \epsffile{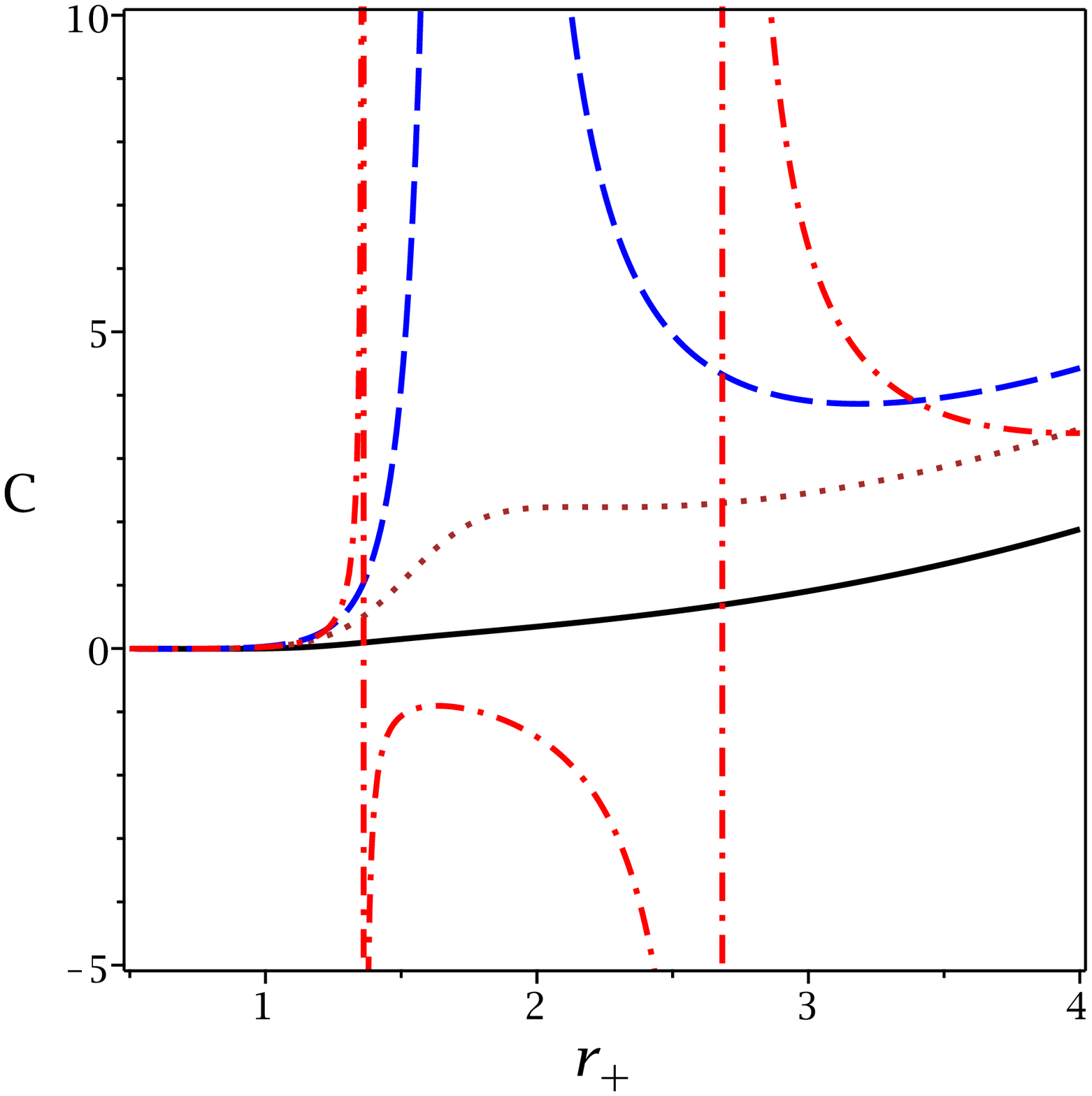}%
\end{array}
$%
\caption{$T$ (left panel) and $C$ (right panel) versus $r_{+}$ for $\Lambda
=-1$, $q=1$, $\protect\alpha =0.5$, $c=c_{1}=c_{2}=c_{3}=2$, $c_{4}=1$, $k=1$%
, $d=5$, $s=0.7$, $m=0$ (continuous line), $m=0.35$ (dotted line), $m=0.4138$
(dashed line) and $m=0.5$ (dashed-dotted line).}
\label{FigC1}
\end{figure}

\begin{figure}[tbp]
$%
\begin{array}{cc}
\epsfxsize=7cm \epsffile{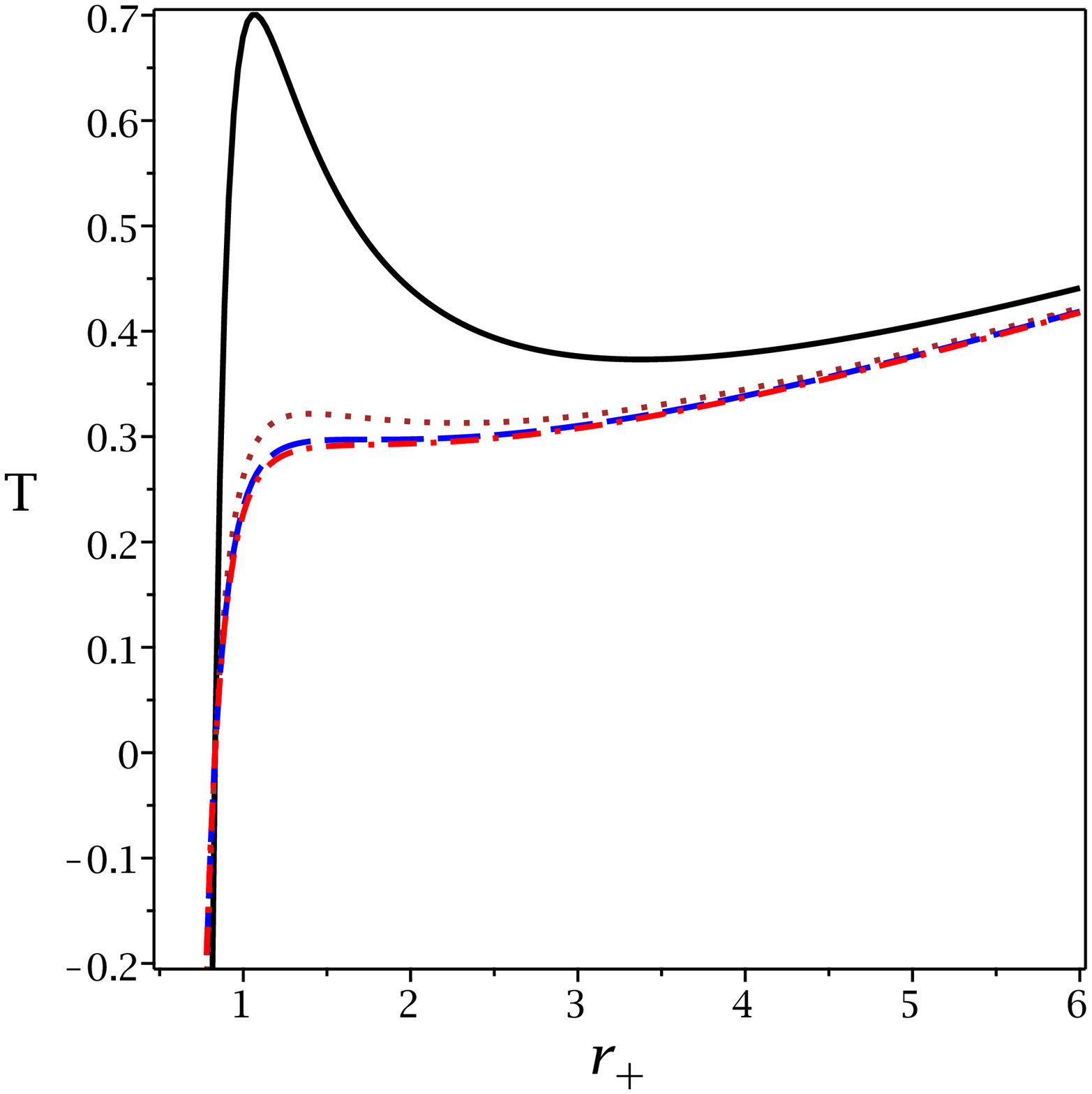} & \epsfxsize=7cm \epsffile{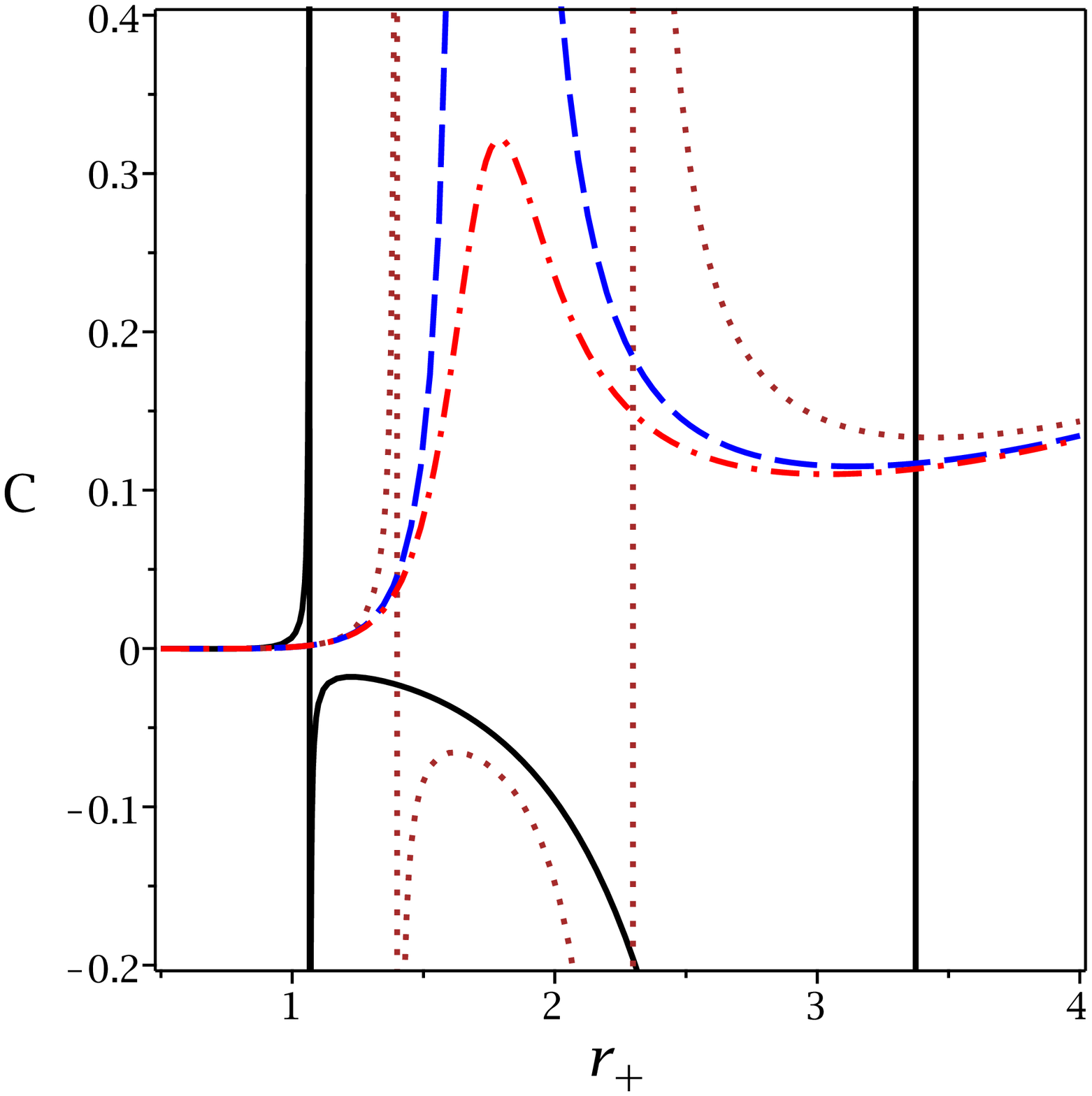}%
\end{array}
$%
\caption{$T$ (left panel) and $C$ (right panel) versus $r_{+}$ for $\Lambda
=-1$, $q=1$, $m=0.4$, $c=c_{1}=c_{2}=c_{3}=2$, $c_{4}=1$, $k=1$, $d=5$, $%
s=0.7$, $\protect\alpha=0$ (continuous line), $\protect\alpha=0.4$ (dotted
line), $\protect\alpha=0.479$ (dashed line) and $\protect\alpha=0.5$
(dashed-dotted line).}
\label{FigC2}
\end{figure}

\begin{figure}[tbp]
$%
\begin{array}{cc}
\epsfxsize=5.5cm \epsffile{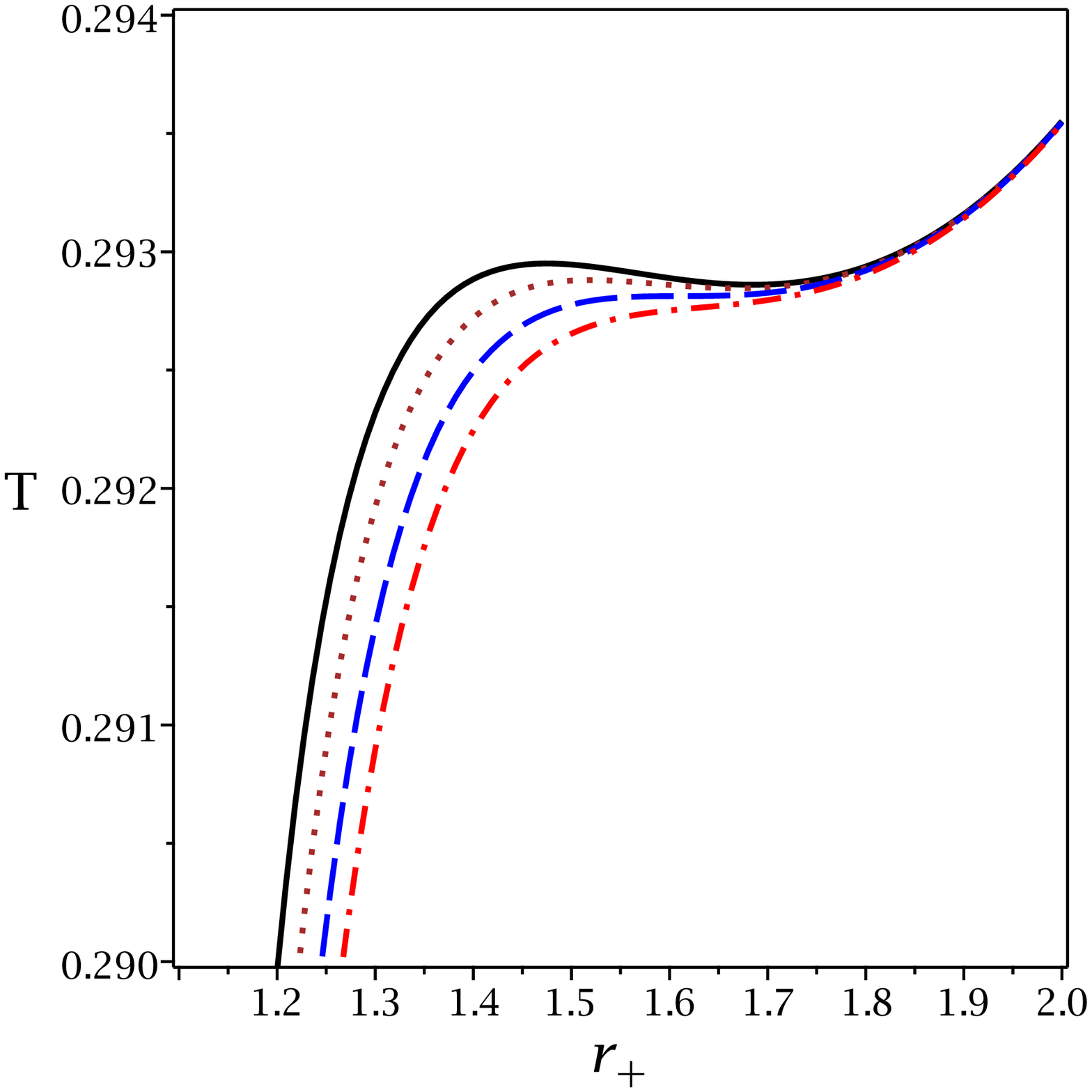} & \epsfxsize=5.5cm \epsffile{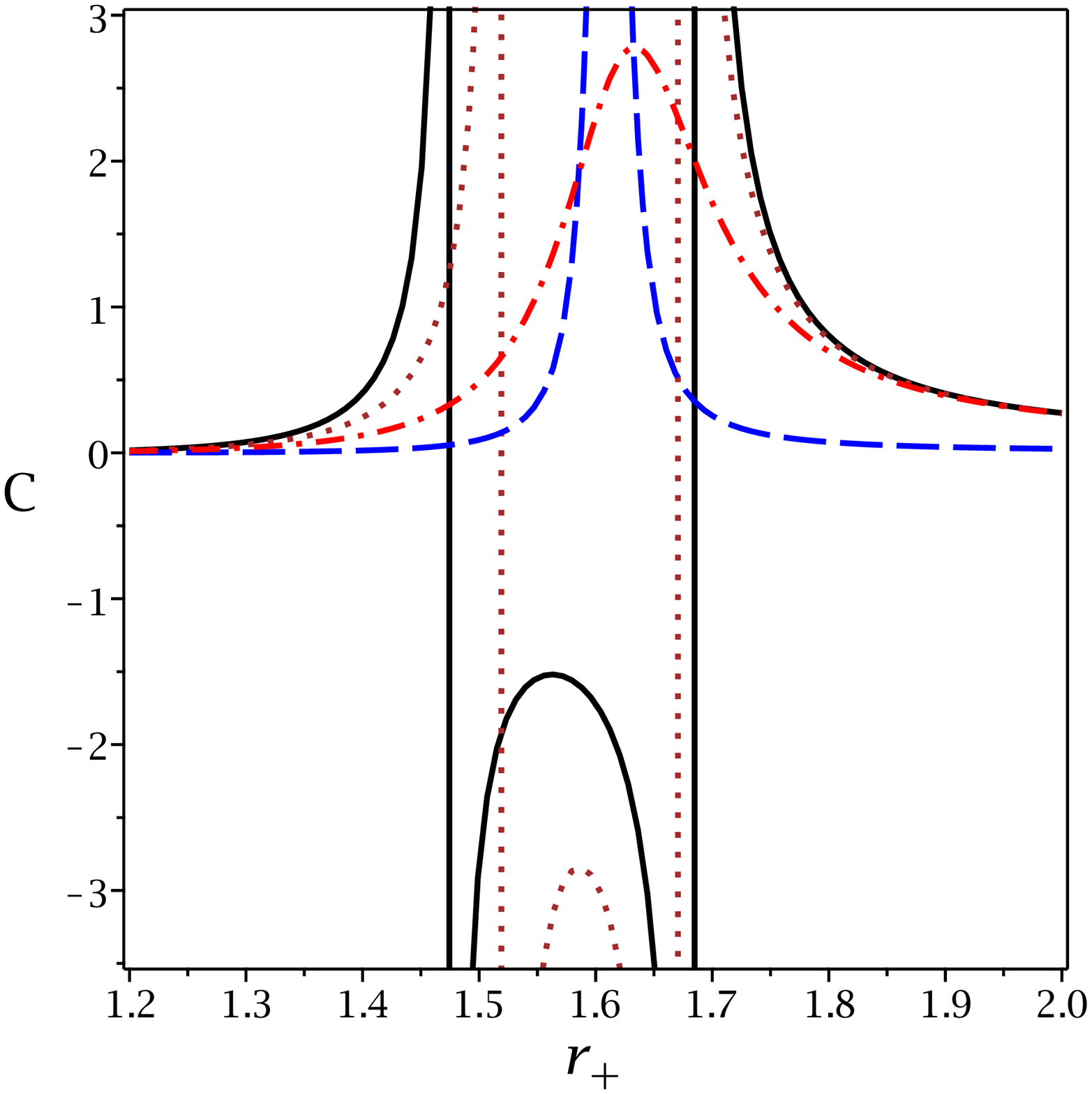}
\\
\epsfxsize=5.5cm \epsffile{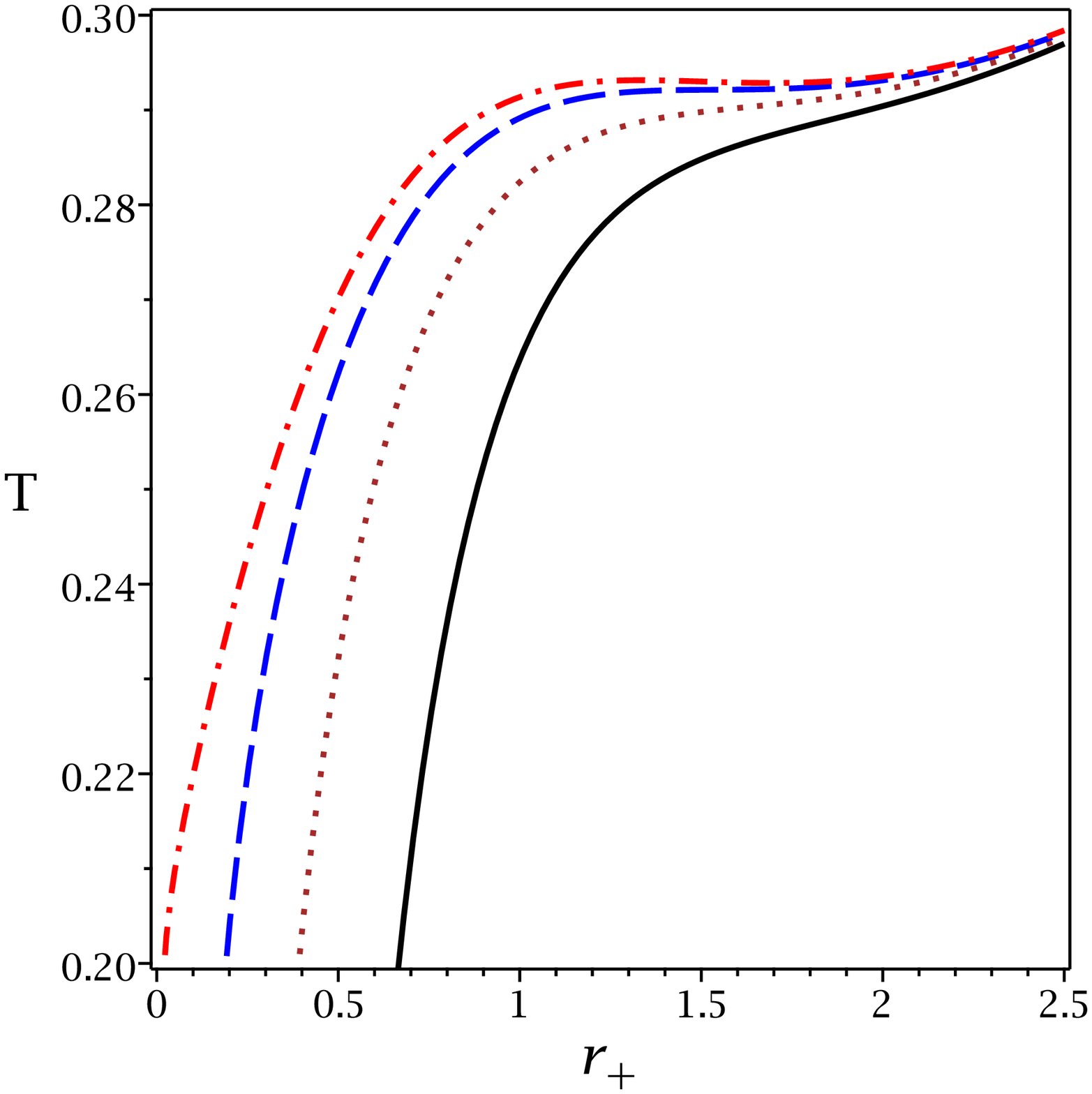} & \epsfxsize=5.5cm \epsffile{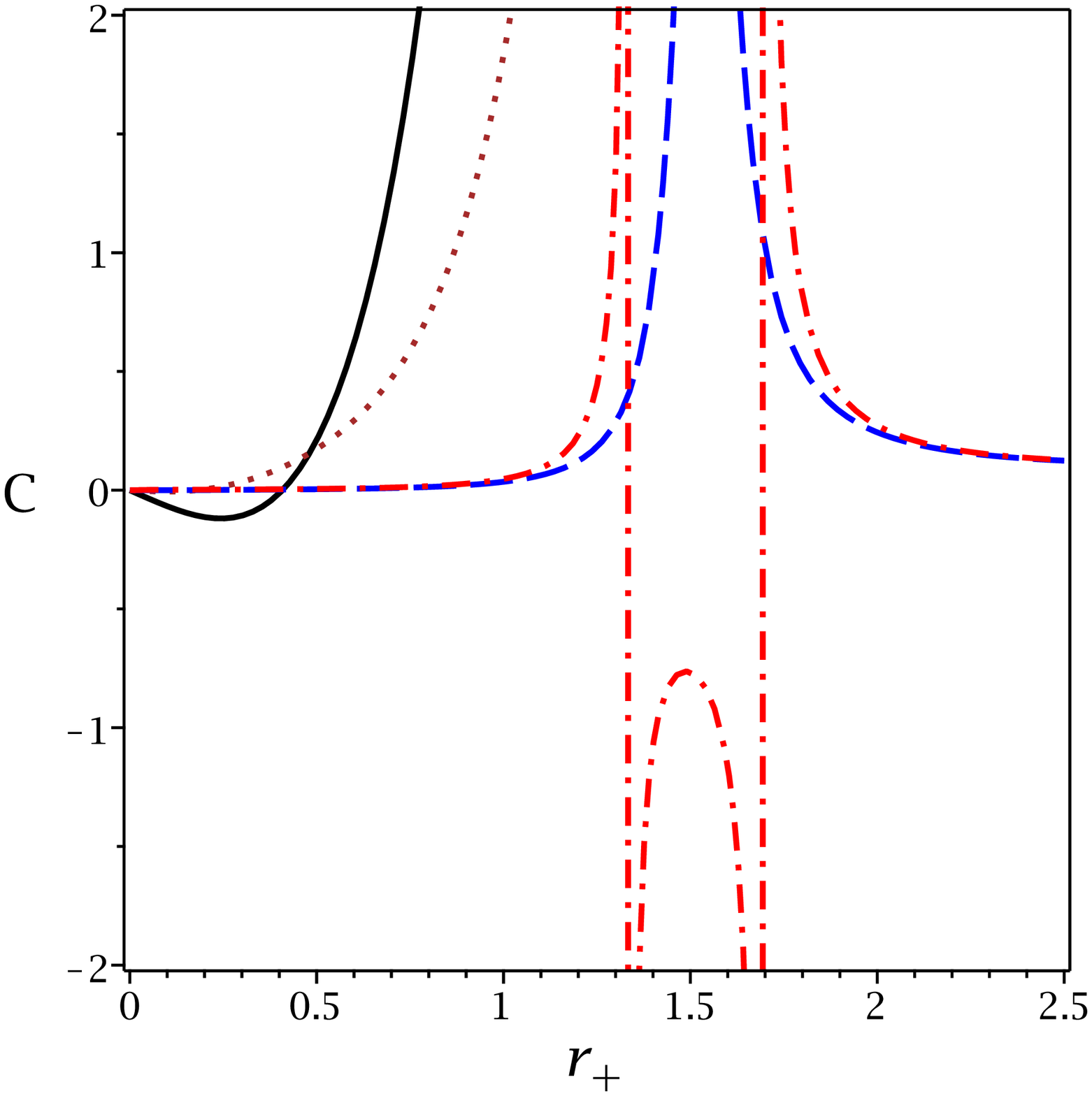}
\\
\epsfxsize=5.5cm \epsffile{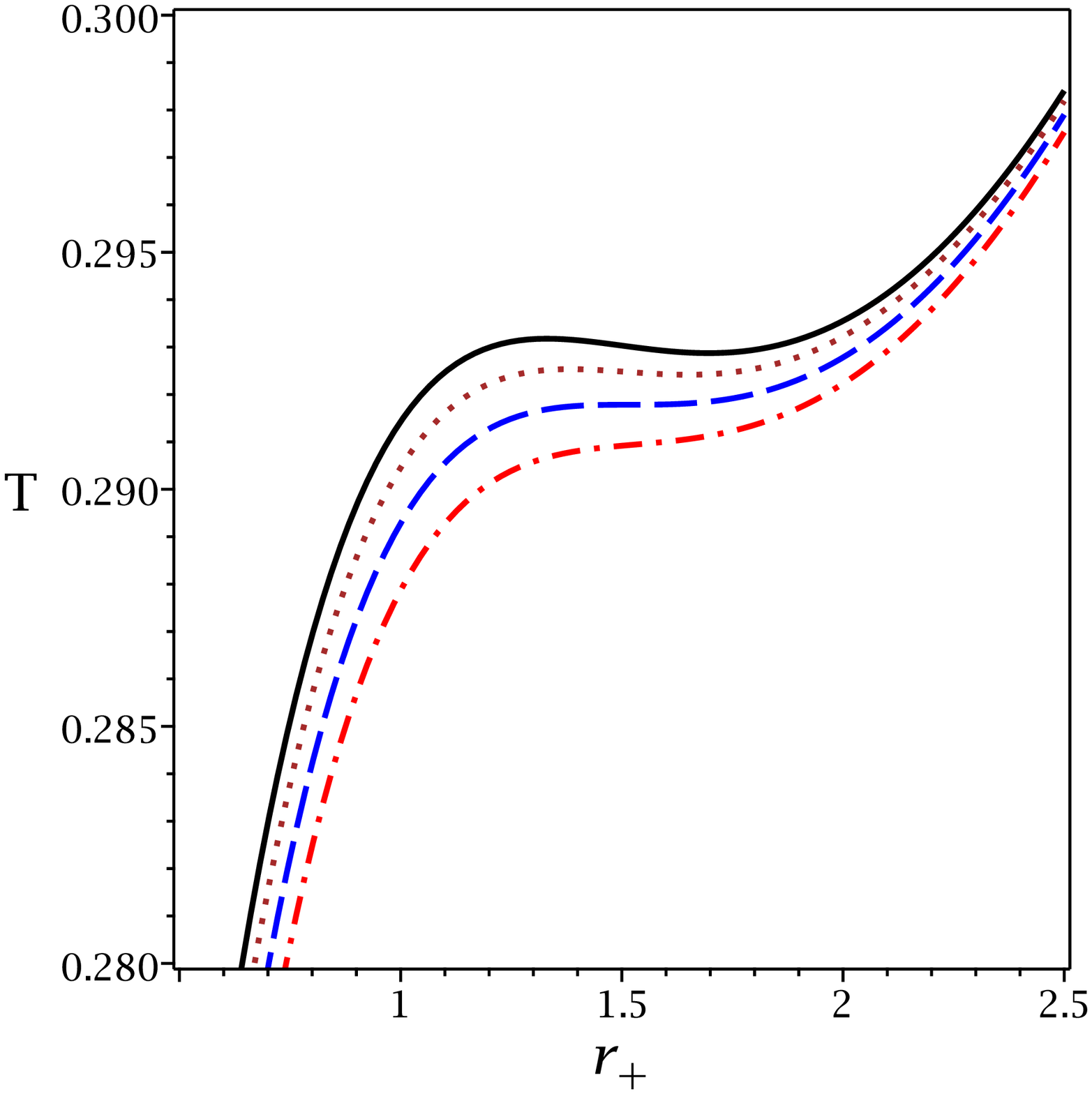} & \epsfxsize=5.5cm \epsffile{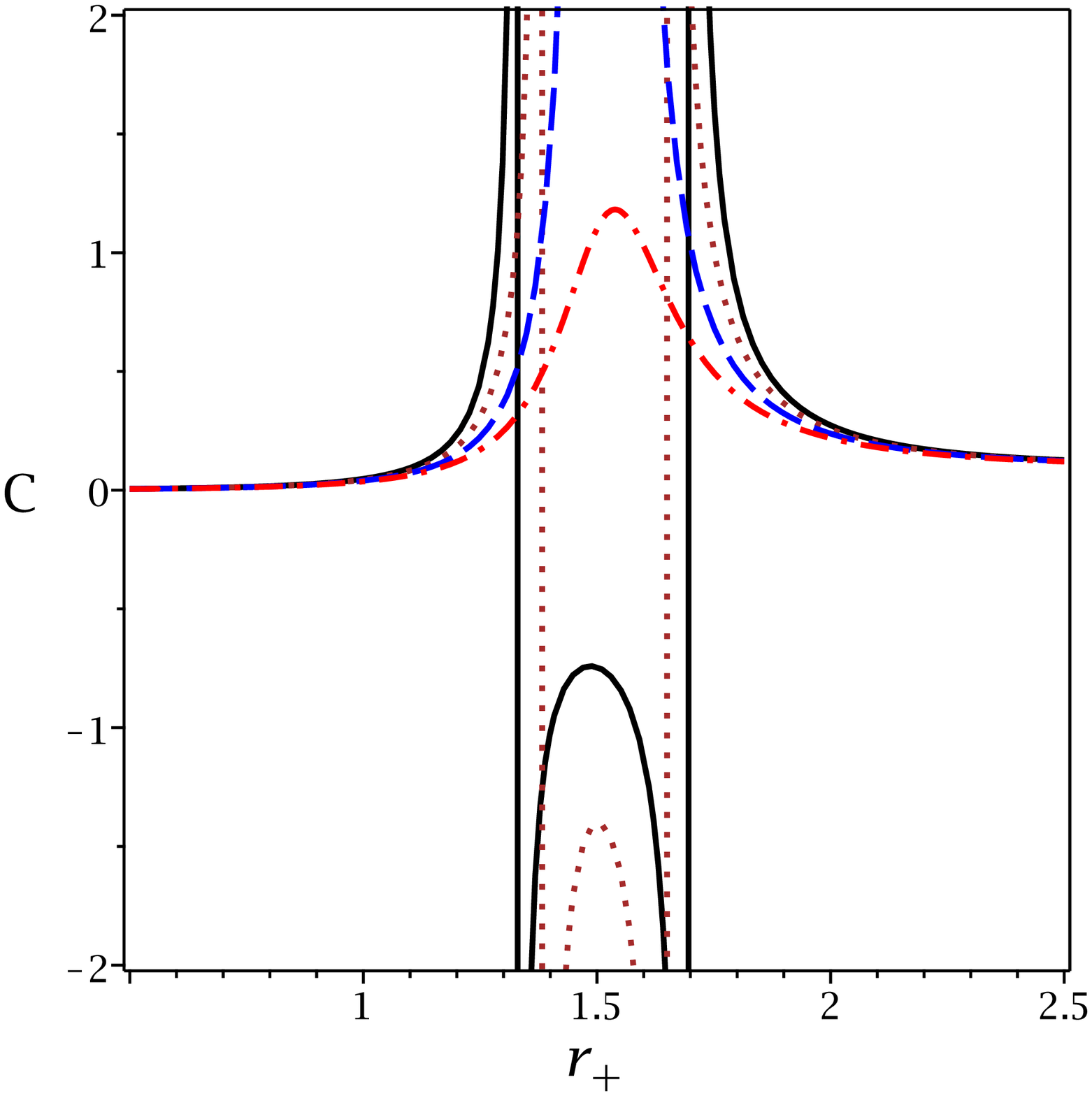}%
\end{array}
$%
\caption{$T$ (left panel) and $C$ (right panel) versus $r_{+}$ for $\Lambda
=-1$, $q=1$, $m=0.4$, $c=c_{1}=c_{2}=c_{3}=2$, $c_{4}=1$, $k=1$, $d=5$, $%
\protect\alpha=0.5$. \newline
Up panels: $s=0.6$ (continuous line), $s=0.61$ (dotted line), $s=0.6205$ (dashed line)
and $s=0.63$ (dashed-dotted line). \newline
Middle panels: $s=1.2$ (continuous line), $s=1.4$ (dotted line), $s=1.56$ (dashed line) and
$s=1.8$ (dashed-dotted line). \newline
Down panels: $s=1.9$ (continuous line), $s=3.2$ (dotted line), $s=3.64$ (dashed line) and
$s=4$ (dashed-dotted line).}
\label{FigC3}
\end{figure}

\begin{figure}[tbp]
$%
\begin{array}{cc}
\epsfxsize=7cm \epsffile{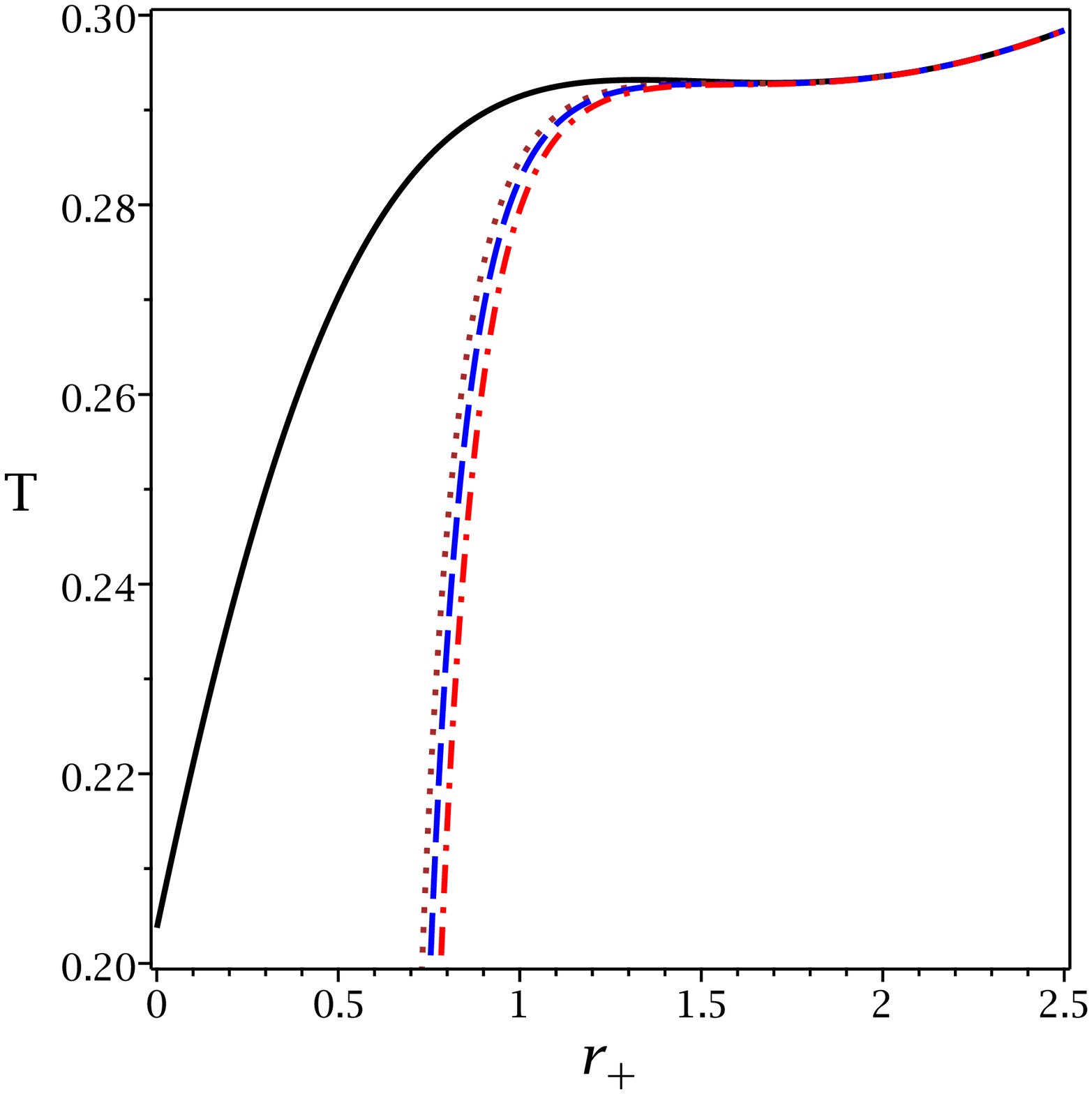} & \epsfxsize=7cm \epsffile{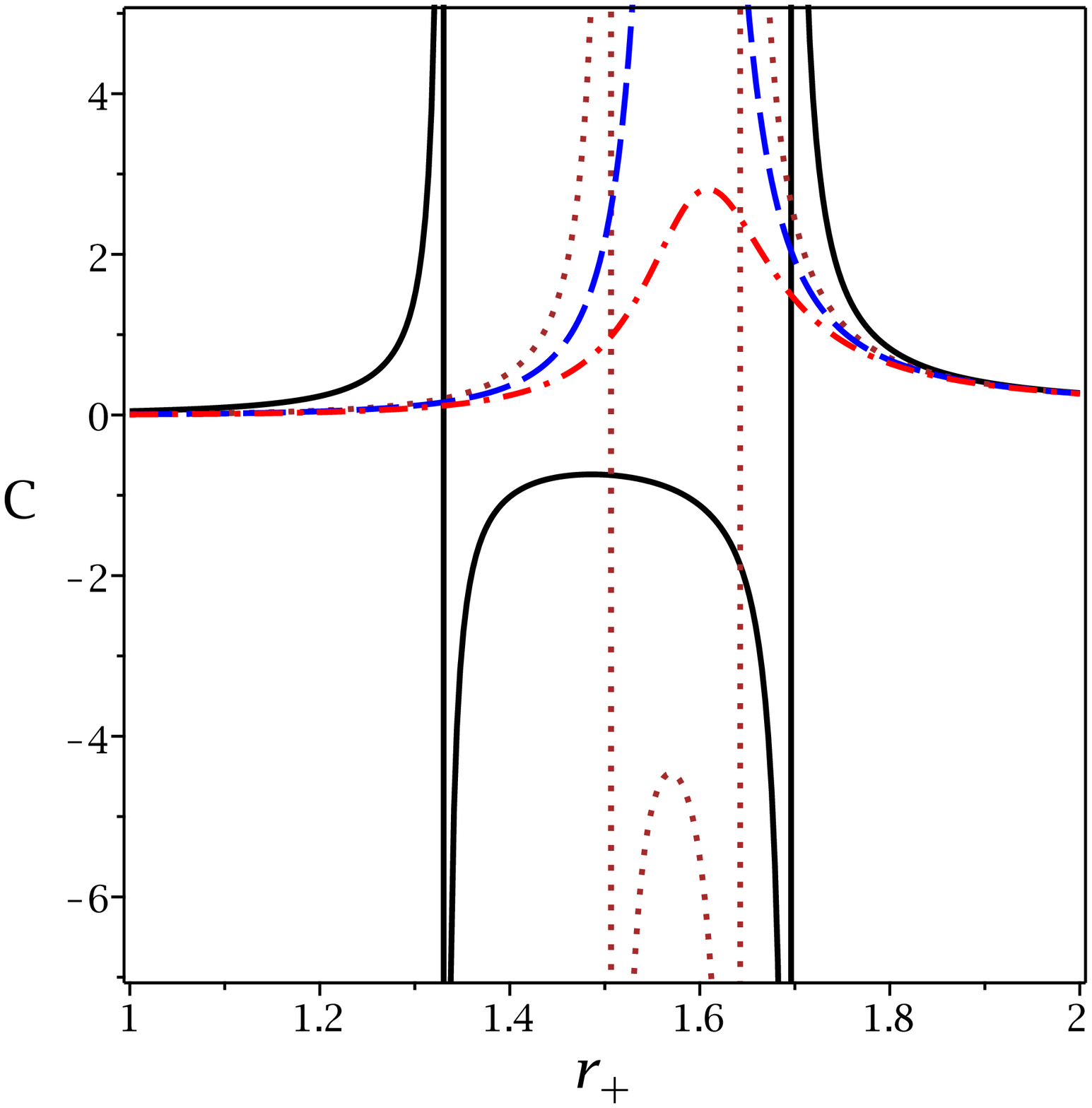}%
\end{array}
$%
\caption{$T$ (left panel) and $C$ (right panel) versus $r_{+}$ for $\Lambda
=-1$, $\protect\alpha=0.5$, $m=0.4$, $c=c_{1}=c_{2}=c_{3}=2$, $c_{4}=1$, $%
k=1 $, $d=5$, $s=0.7$, $q=0$ (continuous line), $q=0.2$ (dotted line), $%
q=0.24$ (dashed line) and $q=0.3$ (dashed-dotted line).}
\label{FigC4}
\end{figure}

\begin{figure}[tbp]
$%
\begin{array}{ccc}
\epsfxsize=5.5cm \epsffile{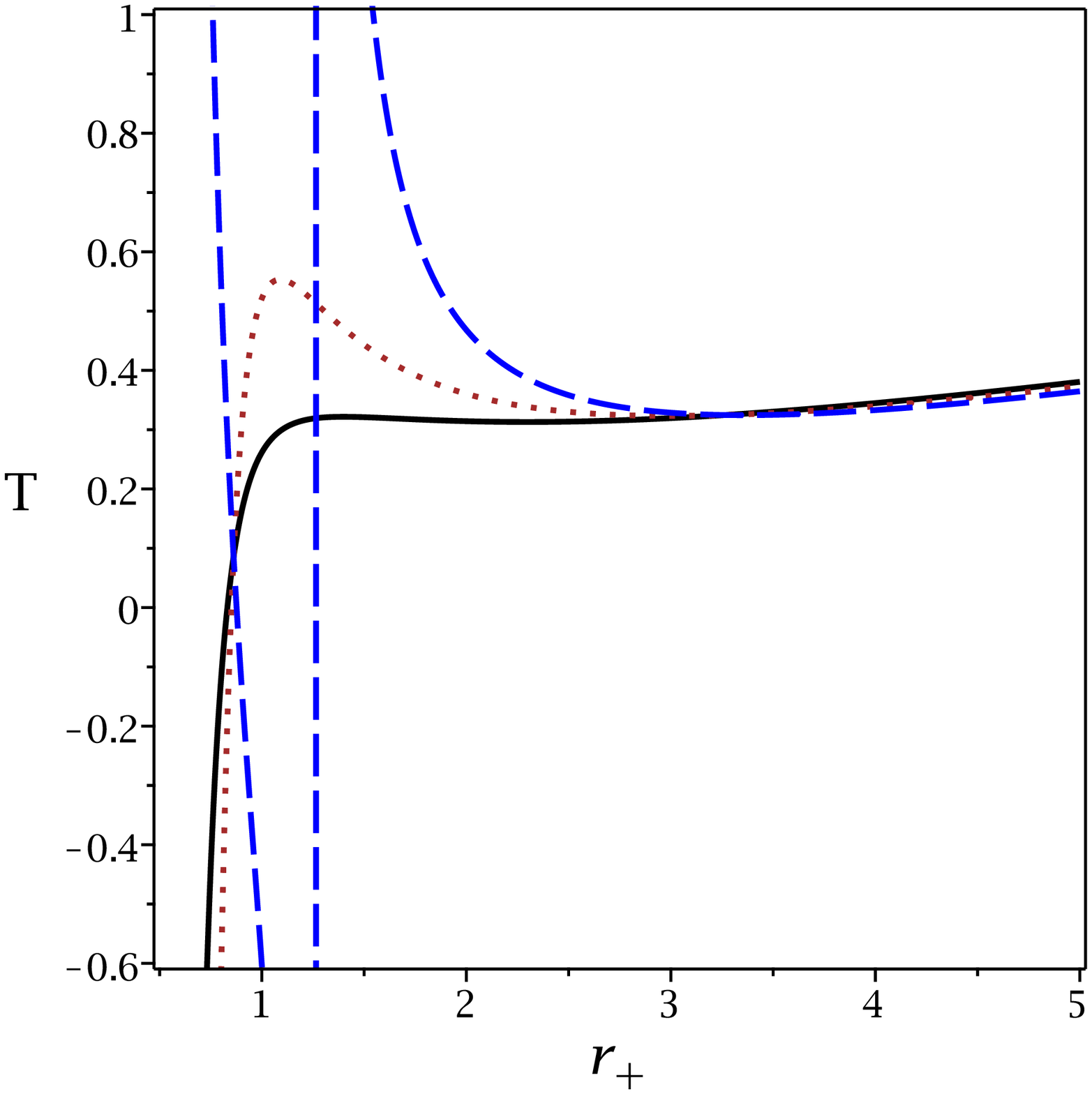} & \epsfxsize=5.5cm \epsffile{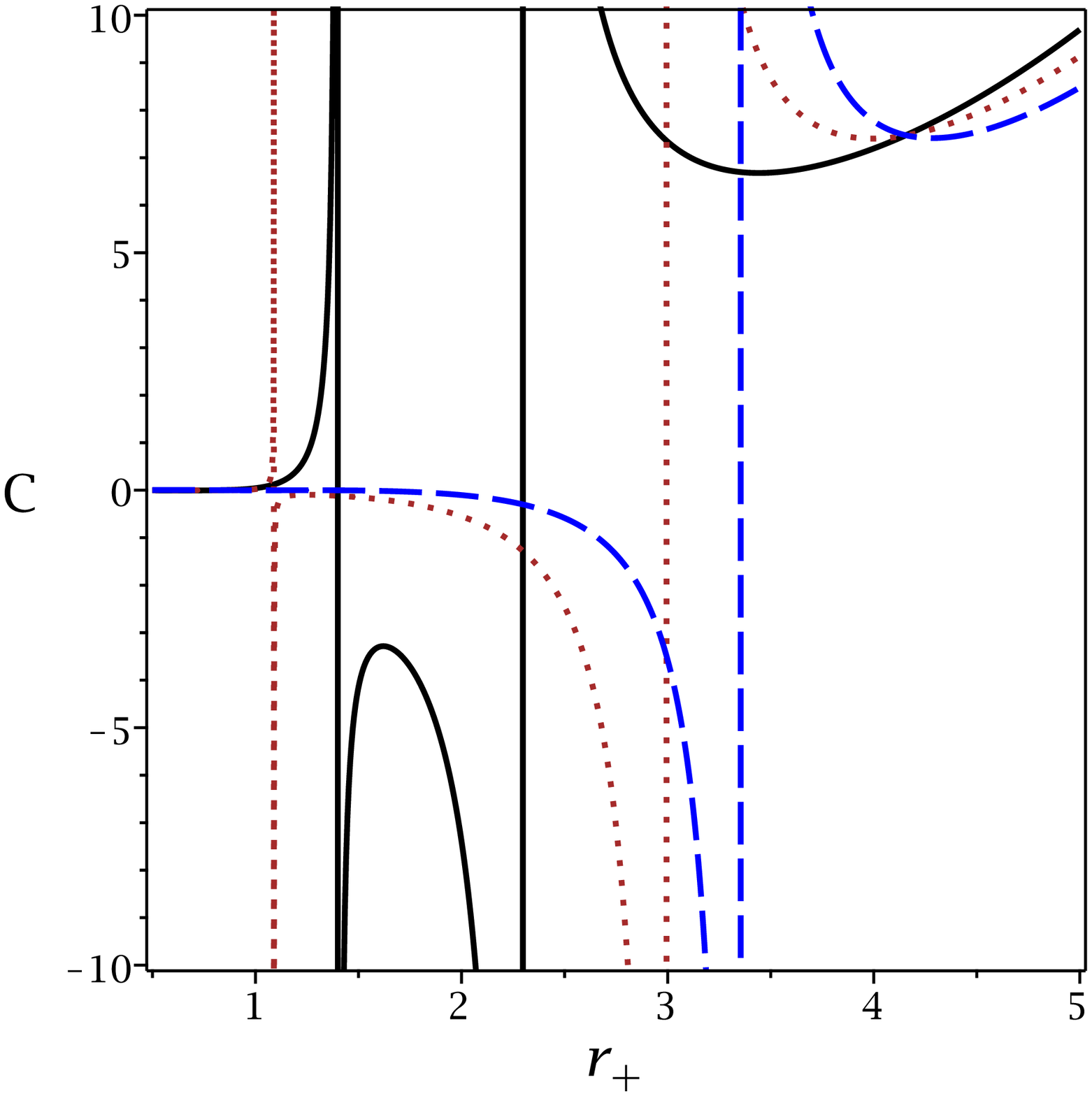} & %
\epsfxsize=5.5cm \epsffile{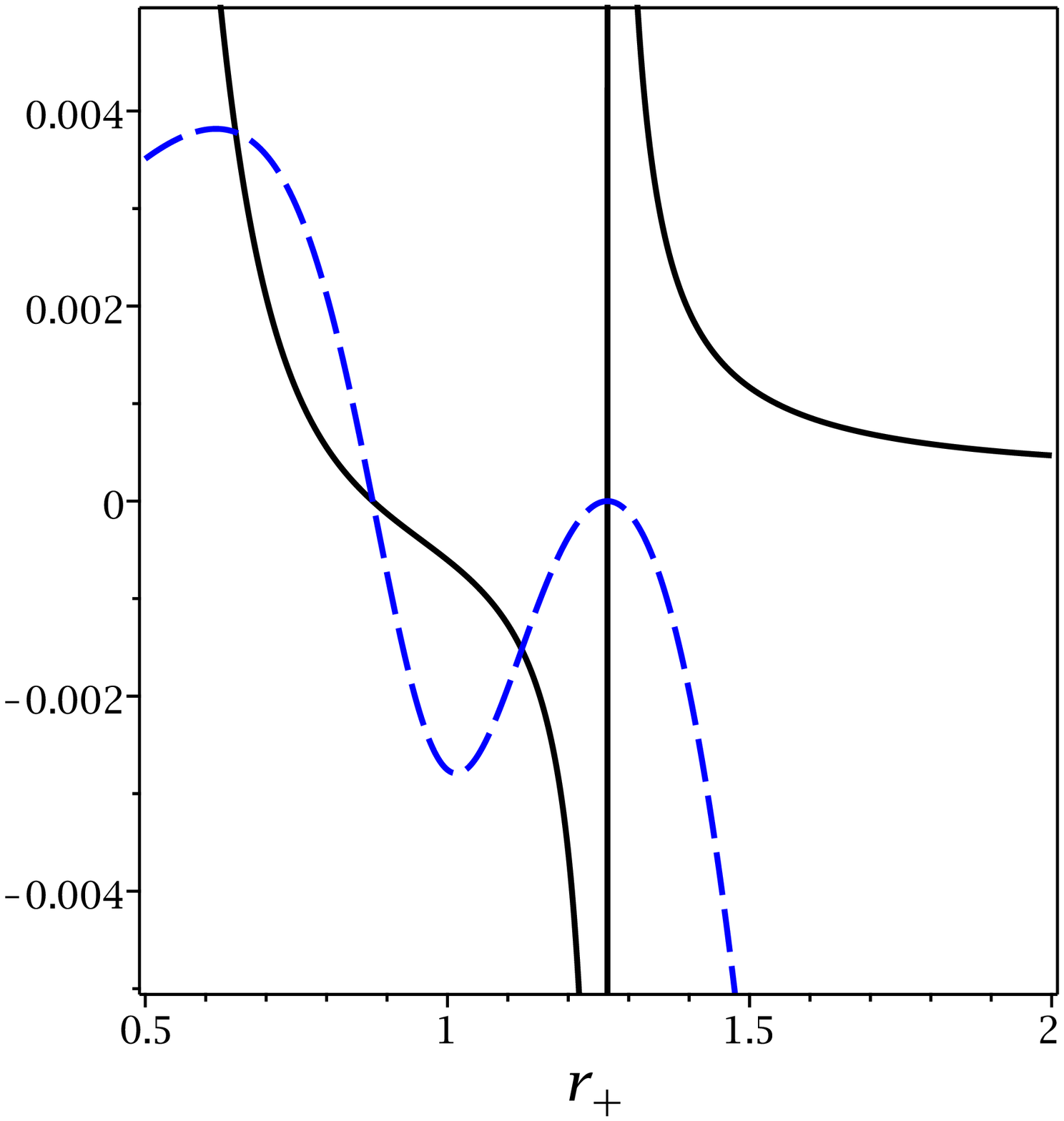}%
\end{array}
$%
\caption{For $\Lambda =-1$, $q=1$, $m=0.4$, $c=c_{1}=c_{2}=c_{3}=2$, $%
c_{4}=1 $, $\protect\alpha=0.4$, $d=5$, $s=0.7$, $k=1$ (continuous line), $%
k=0$ (dotted line) and $k=-1$ (dashed line). \newline
$T$ (left panel) and $C$ (middle panel) versus $r_{+}$. Right panel: $T$
(continuous line) and $C$ (dashed line) versus $r_{+}$ for $k=-1$.}
\label{FigC5}
\end{figure}

Unfortunately, it was not possible to obtain divergence points and roots of
the heat capacity analytically, and therefore, we employed the numerical
method. In order to investigate the effects of different factors on
thermodynamical behavior of the solutions, we have plotted following
diagrams for variation of the gravitons's mass (Fig. \ref{FigC1}), GB factor
(Fig. \ref{FigC2}), nonlinear parameter (Fig. \ref{FigC3}), electric charge
(Fig. \ref{FigC4}) and topological factor (Fig. \ref{FigC5}).

The heat capacity provided in this paper could enjoy one of the following
cases: i) one root and being an increasing function of the horizon radius.
ii) one root with one divergency. iii) one root and two divergencies.
Correspondingly, the temperature enjoys one of the following behaviors: i)
having one root. ii) existence of one root and one extremum. iii) having one
root, one maximum and one minimum. iv) existence of one root, one divergency
and one minimum which is reported only for black holes with hyperbolic
horizon. The root of temperature is coincidence with the root of heat
capacity, except for one case that will be highlighted later. The
divergencies of heat capacity are taking place exactly where temperature
acquires an extremum. Before the root, both heat capacity and
temperature are negative which indicate non-physical unstable solutions. The
divergencies of heat capacity take place only after root. If there is only
one divergency, heat capacity is positive around it, so the solutions
are thermally stable. In this case, our thermodynamical system meets a
critical point at where heat capacity diverges. In the case of two
divergencies, between root and smaller divergence point and after larger
divergency, since heat capacity is positive valued, the solutions are
thermally stable in these regions. Between the divergencies, heat capacity
is negative indicating unstable state. This shows that for this case,
a phase transition takes place between divergencies.

%%%%%%%%%%%%%%%%%%%%%%%%%%%%%%%%%%%%%%%%%%%%%%%%%%%%%%%%%%%%%%%

%%%%%%%%%%%%%%%%%%%%%%%%%%%%%%%%%%%%%%%%%%%%%%%%%%%%%%%%%%%%%%%
%%%%%%%%%%%%%%%%%%%%%%%%%%%%%%%%%%%%%%%%%%%%%%%%%%%%%%%%%%%%%%%

%%%%%%%%%%%%%%%%%%%%%%%%%%%%%%%%%%%%%%%%%%%%%%%%%%%%%%%%%%%%%%%
%%%%%%%%%%%%%%%%%%%%%%%%%%%%%%%%%%%%%%%%%%%%%%%%%%%%%%%%%%%%%%%

%%%%%%%%%%%%%%%%%%%%%%%%%%%%%%%%%%%%%%%%%%%%%%%%%%%%%%%%%%%%%%%
%%%%%%%%%%%%%%%%%%%%%%%%%%%%%%%%%%%%%%%%%%%%%%%%%%%%%%%%%%%%%%%

%%%%%%%%%%%%%%%%%%%%%%%%%%%%%%%%%%%%%%%%%%%%%%%%%%%%%%%%%%%%%%%
%%%%%%%%%%%%%%%%%%%%%%%%%%%%%%%%%%%%%%%%%%%%%%%%%%%%%%%%%%%%%%%

%%%%%%%%%%%%%%%%%%%%%%%%%%%%%%%%%%%%%%%%%%%%%%%%%%%%%%%%%%%%%%%

Evidently, for small values of the massive parameter, the heat capacity
enjoys only one root. The place of root is a decreasing function of this
parameter. Increasing massive gravity results into formation of divergencies
in the structure of heat capacity. Therefore, for sufficiently large values
of the massive parameter, thermodynamical system acquires critical behavior
and phase transition (see Fig. \ref{FigC1}). The effects of GB parameter are
opposite to massive parameter. In other words, by increasing the value of
this parameter, the divergencies of heat capacity, hence phase transitions
in thermal structure of the heat capacity are omitted (Fig. \ref{FigC2}).
The effects of nonlinearity parameter proves to be outstanding. Increasing
this parameter results into periodic behavior for existence of divergencies
and their number. For specific region, increasing the nonlinearity parameter
results into vanishing the divergencies, and heat capacity becomes a smooth
function of the horizon radius (see up panels of Fig. \ref{FigC3}).
Interestingly, increasing the nonlinearity further causes the heat capacity
to acquire the divergencies that were omitted (see middle panels of Fig. \ref%
{FigC3}). Therefore, system will have thermodynamically critical behavior.
But, if the nonlinearity parameter is increased again, the divergencies will
vanish again (see down panels of Fig. \ref{FigC3}). Therefore, we observe a
periodic behavior for existence/absence of thermodynamical phase transition
for the black holes for variation of the nonlinearity parameter. This
specific behavior is solely belongs to this type of the nonlinear
electromagnetic field and Born-Infeld families of the nonlinear
electromagnetic field do not possess such property. The effects of variation
of electric charge is depicted in Fig. \ref{FigC4}. Interestingly, in the
absence of electric charge, the temperature and heat capacity acquires a
non-zero value for vanishing horizon radius (see continuous line in Fig. \ref%
{FigC4}). Considering this specific behavior and evaporation of the black
holes mechanism, one can point out that for these black holes, in
evaporation process (where horizon radius is vanished) a trace of existence
of the black holes is left in term of thermal fluctuation. This is a generic
behavior that is present for the black holes due to massive gravity
generalization. Previously, existence of the trace of thermal fluctuation
for temperature of the black holes in the presence of massive gravity was
reported for $3$ and $4$ dimensional black holes. Presence of it in $5$
dimensional case signals that this is a generic behavior available for black
holes in massive gravity generalization. Existence of such trace could be
used to address the long standing issue of information paradox for black
holes. As we can see, for non-zero value of the electric charge, this
specific behavior is omitted and, temperature and heat capacity acquire a
root. For the absence or small values of the electric charge, heat capacity
enjoys the presence of two divergencies in its structure. Increasing the
value of electric charge results into first one divergency and then its
vanishing. This shows the number of the divergencies is a decreasing
function of the electric charge. Finally, as one can see, the topological
factor, modifies the place of divergencies in heat capacity and their number
as well (Fig. \ref{FigC5}). One of the most interesting results is regarding
black holes with hyperbolic horizon. For this specific horizon, temperature
enjoys a root, a divergency and one minimum in its structure. Before the
root and after the divergency, temperature is positive valued whereas
between root and divergency, temperature is negative, therefore solutions
are nonphysical. Interestingly, the divergency of temperature is coincided
with an extremum root in the heat capacity (see right panel of Fig. \ref%
{FigC5}). For $k=-1$, the root of temperature coincides with a non-extremum root in the
heat capacity and extremum of the temperature is matched with a divergency
in the heat capacity. Before the root of temperature, both temperature and
heat capacity are positive valued indicating existence of physical stable
state. Between the root and divergency of the temperature, both heat
capacity and temperature are negative valued, hence solutions are not
physical. Between divergency of the temperature and its minimum, temperature
is positive whereas, interestingly, heat capacity is negative valued
indicating the existence of physical unstable state for black holes. After
the minimum of temperature (divergency of heat capacity), both temperature
and heat capacity are positive. This shows that there is a phase transition
between medium unstable black holes and larger stable ones. If one compare
the hyperbolic case with spherical/flat black holes, one can see that
thermodynamically speaking, there are significant differences between the
thermodynamical structure and phase transitions of black holes in these two
horizons. On the other hand, as we can see, not all the roots in heat
capacity are matched with roots in the temperature. Interestingly, as we
study the heat capacity, no discontinuity is observed where temperature is
divergent. This shows the necessity of investigating temperature and heat
capacity separately and also together to draw a better picture regarding the
thermodynamical behavior of the black holes.

The nonzero massive parameter is related to the mass of gravitons.
As we can see, the increment in this parameter results into system
acquiring a critical thermodynamical behavior. Depending on the
values of $m$, system may acquire specific critical behavior.
Recently, it was pointed out that quasi normal modes of the black
holes present a characteristic behavior in their frequencies which
exactly takes place where system has a thermodynamical phase
transition. Considering this fact, the specific contributions of
the massive gravity into thermodynamical structure of the black
holes could be employed to distinguish the validity of massive
gravity generalization. As for the GB gravity, studying the Ricci
scalar or Riemann tensor shows that the effects of GB gravity is
toward increasing the value of these quantities. Increment in the
value of curvature scalar indicates stronger gravitational power.
Therefore, on can conclude that strength of the gravitational
field is an increasing function of the GB parameter. Considering
this fact, we can see that by increasing the value of GB
parameter, hence the strong gravitational field, the critical
behavior is omitted. This shows that the presence of an
instability, hence a phase transition, is more probable to be seen
in black holes with weaker gravitational field, the existence of a
phase transition is less possible or even in some cases impossible.
This shows that as we increase the strength of gravitational
field, a rearrange in thermodynamical distribution of the system
is done in a manner to avoid thermal phase transition. This also
shows that gravitational strength and, thermodynamical behavior
and properties are closely related and the modification in one has
more direct and observable effects on the other one. The
topological factor represents the geometrical structure of the
horizon for black holes provided in this paper. As we have pointed
out, depending on the type of horizon, thermodynamical structure
of the black holes would be significantly different. This signals
us with the fact that geometrical structure of our object of
interest also plays a crucial role in determining the
thermodynamical behavior of it. Therefore, one can conclude that
there is also a direct relation between thermodynamical properties
of the system (including phase transitions), and geometrical
structure. In essence, we are expecting this fact. The reason is
because through Einstein point of the view of gravity, gravity is
induced due to geometrical variation of the spacetime. Such
variation is originated from the effects of mass on the spacetime.
Therefore, there is a direct relation between gravitational field
and geometrical properties. Establishing the fact that
gravitational field directly affects the thermodynamical behavior,
one can conclude that there is a direct relation between
geometrical structure and thermodynamical properties. In our study
of the heat capacity, such link between geometrical properties and
thermodynamical behavior was highlighted. This signals us with the
fact that the origin of thermodynamical behavior, geometrical
structure and gravitational field could be the same. In the
context of electric charge, we observed that increasing electric
charge results into omitting the divergencies in the heat
capacity. One can conclude that for super charged black holes,
system is restricted in a manner that prevents the thermodynamical
structure reaching critical behavior. This shows that the total
net charge provided for black holes could modify the structure of
black holes to a level of canceling the effects of the massive
gravity and GB generalization. Comparing to other generalizations,
the electric charge is of a more free nature. This is due to fact
that black holes by consuming the matters surrounding them, can
achieve higher values of the electric charge. This shows that
comparing to other parameters discussed in the paper, the electric
charge could grow more rapidly comparing to other factors
considered and discussed before. Once more, we point out that in
the absence of electric field, the generalization to massive
gravity provided black holes with a possibility of existence of
non-zero temperature and heat capacity for vanishing horizon
radius. This indicates the presence of a trace of existence of
black holes even after black hole evaporation. This is a generic
property which so far was observed only for black holes in the
presence of massive gravity. Finally, the periodic effect of
nonlinearity parameter on the existence of divergencies in heat
capacity, hence phase transition should be pointed out. The
nonlinearity parameter measures the degree of nonlinear nature of
the electromagnetic field. In the context of Born-Infeld family of
nonlinear electromagnetic fields, increasing the nonlinearity
parameter results into decreasing the nonlinear nature of the
electromagnetic field. In other words, for large values of the
nonlinearity parameter, electromagnetic field in Born-Infeld
family will yield a Maxwell-like behavior. Interestingly, for PMI
case, the Maxwell-like case is achieved for only a specific case
($s=1$) and variation of the nonlinearity parameter results into
periodic behavior regarding the existence of phase transition
points. This shows that PMI theory enjoys the larger number of the
possibilities regarding existence of the critical behavior
comparing to Born-Infeld theories.

\begin{acknowledgements}
We thank Shiraz University Research Council. This work has been
supported financially by the Research Institute for Astronomy and
Astrophysics of Maragha, Iran.
\end{acknowledgements}

\end{document}